\PassOptionsToPackage{unicode}{hyperref}
\PassOptionsToPackage{hyphens}{url}
\PassOptionsToPackage{dvipsnames,svgnames,x11names}{xcolor}
\documentclass[
]{article}
\usepackage{xcolor}
\usepackage[left=1in,right=1in,bottom=1in]{geometry}
\usepackage{amsmath,amssymb}
\usepackage{iftex}
\ifPDFTeX
  \usepackage[T1]{fontenc}
  \usepackage[utf8]{inputenc}
  \usepackage{textcomp} 
\else 
  \usepackage{unicode-math} 
  \defaultfontfeatures{Scale=MatchLowercase}
  \defaultfontfeatures[\rmfamily]{Ligatures=TeX,Scale=1}
\fi
\usepackage{lmodern}
\ifPDFTeX\else
\fi
\IfFileExists{upquote.sty}{\usepackage{upquote}}{}
\IfFileExists{microtype.sty}{
  \usepackage[]{microtype}
  \UseMicrotypeSet[protrusion]{basicmath} 
}{}
\makeatletter
\@ifundefined{KOMAClassName}{
  \IfFileExists{parskip.sty}{%
    \usepackage{parskip}
  }{
    \setlength{\parindent}{0pt}
    \setlength{\parskip}{6pt plus 2pt minus 1pt}}
}{
  \KOMAoptions{parskip=half}}
\makeatother
\makeatletter
\ifx\paragraph\undefined\else
  \let\oldparagraph\paragraph
  \renewcommand{\paragraph}{
    \@ifstar
      \xxxParagraphStar
      \xxxParagraphNoStar
  }
  \newcommand{\xxxParagraphStar}[1]{\oldparagraph*{#1}\mbox{}}
  \newcommand{\xxxParagraphNoStar}[1]{\oldparagraph{#1}\mbox{}}
\fi
\ifx\subparagraph\undefined\else
  \let\oldsubparagraph\subparagraph
  \renewcommand{\subparagraph}{
    \@ifstar
      \xxxSubParagraphStar
      \xxxSubParagraphNoStar
  }
  \newcommand{\xxxSubParagraphStar}[1]{\oldsubparagraph*{#1}\mbox{}}
  \newcommand{\xxxSubParagraphNoStar}[1]{\oldsubparagraph{#1}\mbox{}}
\fi
\makeatother

\usepackage{longtable,booktabs,array}
\usepackage{calc} 
\usepackage{etoolbox}
\makeatletter
\patchcmd\longtable{\par}{\if@noskipsec\mbox{}\fi\par}{}{}
\makeatother
\IfFileExists{footnotehyper.sty}{\usepackage{footnotehyper}}{\usepackage{footnote}}
\makesavenoteenv{longtable}
\usepackage{graphicx}
\makeatletter
\newsavebox\pandoc@box
\newcommand*\pandocbounded[1]{
  \sbox\pandoc@box{#1}%
  \Gscale@div\@tempa{\textheight}{\dimexpr\ht\pandoc@box+\dp\pandoc@box\relax}%
  \Gscale@div\@tempb{\linewidth}{\wd\pandoc@box}%
  \ifdim\@tempb\p@<\@tempa\p@\let\@tempa\@tempb\fi
  \ifdim\@tempa\p@<\p@\scalebox{\@tempa}{\usebox\pandoc@box}%
  \else\usebox{\pandoc@box}%
  \fi%
}
\def\fps@figure{htbp}
\makeatother

\NewDocumentCommand\citeproctext{}{}

\makeatletter
 \let\@cite@ofmt\@firstofone
 \def\@biblabel#1{}
 \def\@cite#1#2{{#1\if@tempswa , #2\fi}}
\makeatother
\newlength{\cslhangindent}
\newlength{\csllabelwidth}
\newenvironment{CSLReferences}[2] 
 {\begin{list}{}{%
  \setlength{\itemindent}{0pt}
  \setlength{\leftmargin}{0pt}
  \setlength{\parsep}{0pt}
  \ifodd #1
   \setlength{\leftmargin}{\cslhangindent}
   \setlength{\itemindent}{-1\cslhangindent}
  \fi
  \setlength{\itemsep}{#2\baselineskip}}}
 {\end{list}}
\usepackage{calc}

\providecommand{\tightlist}{%
  \setlength{\itemsep}{0pt}\setlength{\parskip}{0pt}}

\usepackage{booktabs}
\usepackage{longtable}
\usepackage{array}
\usepackage{multirow}
\usepackage{wrapfig}
\usepackage{float}
\usepackage{colortbl}
\usepackage{pdflscape}
\usepackage{tabu}
\usepackage{threeparttable}
\usepackage{threeparttablex}
\usepackage[normalem]{ulem}
\usepackage{makecell}
\usepackage{xcolor}
\usepackage{placeins}
\makeatletter
\@ifpackageloaded{float}{}{\usepackage{float}}
\floatstyle{plain}
\@ifundefined{c@chapter}{\newfloat{suppfig}{h}{losfig}}{\newfloat{suppfig}{h}{losfig}[chapter]}
\floatname{suppfig}{S-Figure}

\makeatother
\makeatletter
\@ifpackageloaded{caption}{}{\usepackage{caption}}
\AtBeginDocument{%
\ifdefined\contentsname
  \renewcommand*\contentsname{Table of contents}
\else
  \newcommand\contentsname{Table of contents}
\fi
\ifdefined\listfigurename
  \renewcommand*\listfigurename{List of Figures}
\else
  \newcommand\listfigurename{List of Figures}
\fi
\ifdefined\listtablename
  \renewcommand*\listtablename{List of Tables}
\else
  \newcommand\listtablename{List of Tables}
\fi
\ifdefined\figurename
  \renewcommand*\figurename{Figure}
\else
  \newcommand\figurename{Figure}
\fi
\ifdefined\tablename
  \renewcommand*\tablename{Table}
\else
  \newcommand\tablename{Table}
\fi
}
\@ifpackageloaded{float}{}{\usepackage{float}}
\floatstyle{ruled}
\@ifundefined{c@chapter}{\newfloat{codelisting}{h}{lop}}{\newfloat{codelisting}{h}{lop}[chapter]}
\floatname{codelisting}{Listing}

\makeatother
\makeatletter
\usepackage{pdflscape}
\makeatother
\makeatletter
\@ifpackageloaded{caption}{}{\usepackage{caption}}
\@ifpackageloaded{subcaption}{}{\usepackage{subcaption}}
\makeatother
\usepackage{bookmark}
\IfFileExists{xurl.sty}{\usepackage{xurl}}{} 
\makeatletter
\@ifundefined{xmpquote}{\newcommand{\xmpquote}[1]{#1}}{}
\makeatother
\hypersetup{
  pdftitle={Estimating Stability of Individual-Specific Predictions from AI-based prediction models using Bootstrap and Cross-validation Resampling},
  pdfauthor={Alex Carriero; Martijn L.G. van Dam; Anne A.H. de Hond; Richard D. Riley; Gary S. Collins; Karel G.M. Moons; Maarten van Smeden; Nicole S. Erler},
  pdfkeywords={\xmpquote{Bootstrap}, \xmpquote{Clinical prediction
models}, \xmpquote{Cross-validation}, \xmpquote{Prediction
instability}, \xmpquote{Resampling}},
  colorlinks=true,
  linkcolor={blue},
  filecolor={Maroon},
  citecolor={Blue},
  urlcolor={Blue},
  pdfcreator={LaTeX via pandoc}}

\title{Estimating Stability of Individual-Specific Predictions from
AI-based prediction models using Bootstrap and Cross-validation
Resampling}
\author{Alex Carriero\textsuperscript{1,*} \and Martijn L.G. van
Dam\textsuperscript{1,*} \and Anne A.H. de
Hond\textsuperscript{1} \and Richard D.
Riley\textsuperscript{2,3} \and Gary S.
Collins\textsuperscript{2,3} \and Karel G.M.
Moons\textsuperscript{1} \and Maarten van
Smeden\textsuperscript{1} \and Nicole S. Erler\textsuperscript{1}}
\date{}
\begin{document}
\maketitle
\begin{center}
\small
\textsuperscript{1}Julius Center for Health Sciences and Primary
Care, University Medical Center Utrecht, Utrecht, The Netherlands\par
\textsuperscript{2}Department of Applied Health Sciences, University of
Birmingham, Birmingham, United Kingdom\par
\textsuperscript{3}National Institute for Health and Care Research
(NIHR) Biomedical Research Centre, Birmingham, United Kingdom.\par
\def\equalcontributionnote{\par\smallskip * These authors contributed equally to this work.\par}
\def\equalcontributionnote{\par\smallskip * These authors contributed equally to this work.\par}
\ifdefined\equalcontributionnote\equalcontributionnote\fi
\end{center}
\begin{abstract}
Clinical prediction models are typically developed in a single finite
dataset, which is one of many possible datasets that could have been
collected from the target population. Applying the same
model-development procedure in another sample will provide different
predictions for an individual. The degree to which an individual's
prediction varies across development sets is referred to as
\emph{prediction instability} and is not captured by average performance
measures. Resampling of the training data offers a practical way to
assess instability, but its ability to do so accurately and reliably for
different model choices is unclear. In a simulation study, we considered
non-parametric, parametric, and out-of-bag bootstrap methods and
repeated 10-, 5-, and 2-fold cross-validation to estimate metrics of
prediction instability for Elastic Net logistic regression and Random
Forest. The same metrics estimated in independent development samples
provided the reference target. We assessed individual-level prediction
variance, empirical 95\% range width, and mean absolute differences from
the original prediction and true outcome probability. Instability
increased in smaller development samples, was generally greater for
Random Forest, and varied across the probability range. Repeated 5- and
10-fold cross-validation generally underestimated prediction instability
across scenarios. Default non-parametric bootstrap more closely
approximated independent sampling for Elastic Net but often
overestimated instability for Random Forest, as did parametric
bootstrap. Out-of-bag bootstrap and repeated 2-fold cross-validation
generally agreed most closely with independent sampling, although no
procedure was uniformly best. Resampling procedures should not be used
interchangeably, nor should less variable predictions be taken as
evidence of greater stability. To estimate prediction instability,
prefer procedures that predict on held-out data and limit overlap
between training resamples.
\end{abstract}


\section{Introduction}\label{introduction}

Clinical prediction models provide estimates of an individual's
probability of an outcome (Hendriksen et al. 2013; Smeden et al. 2021).
These estimates may guide treatment and monitoring decisions and support
communication with patients. However, the estimated probability is
usually a single value and gives no indication of how much uncertainty
surrounds this estimate. This uncertainty matters when a model is
critically appraised, when decisions are based on probability
thresholds, and when probability estimates are communicated (Riley et
al. 2025).

One source of uncertainty is the finite sample used to develop the
model. Applying the same model-development procedure to another
independent sample of the same size from the same population will
produce a different model and, consequently, a different probability
estimate for the same individual. We refer to this sampling-related
variation as \emph{prediction instability} (Riley and Collins 2023;
Altman and Andersen 1989). Its magnitude can differ between individuals
because some predictor profiles are supported by more development data
than others, and it generally increases as the development sample
becomes smaller (Thomassen et al. 2024; Pate et al. 2020). A model may
therefore have acceptable population-level discrimination and
calibration while still producing unstable predictions for individuals
(Pate et al. 2020).

Ideally, the individual-specific distribution of probability estimates
would be obtained by repeating the complete model-development procedure
in infinitely many independent development samples. In practice, only
one finite development sample is usually available, so the distribution
must be approximated using a finite number of resamples. To assess
prediction instability, Riley and Collins (2023) recommend drawing
same-sized non-parametric bootstrap samples from the development data,
repeating all data-driven modelling steps in each sample, and examining
the resulting predictions. This recommendation is also included in the
TRIPOD+AI guidance for reporting prediction models that use regression
or machine learning methods (Collins et al. 2024). They also call for
research comparing the bootstrap with other resampling approaches. It
remains unclear how closely the bootstrap approximates the independent
sampling distribution and whether this depends on the model-development
procedure.

Other established resampling methods may offer alternative
approximations to the independent sampling distribution, but they do not
necessarily target the same distribution. Non-parametric bootstrap,
parametric bootstrap, out-of-bag bootstrap, and repeated
cross-validation differ in how observations are resampled, how much of
the observed data is used for model fitting, and whether the model used
to predict for an individual was fitted using that individual's data.
These differences may be particularly relevant when tuning parameters
are estimated from the data or when tree-based algorithms are used
(Riley et al. 2021; Barreñada et al. 2024). A recent empirical study
found that estimates of prediction instability differed between the
non-parametric bootstrap and repeated 5-fold cross-validation and
between logistic regression and random forest (Khongthon et al. 2026).
However, the methods were not compared against a known independent
sampling distribution.

In this simulation study, we investigate how well practical resampling
methods estimate prediction instability. We use repeated independent
sampling as the reference and compare the non-parametric bootstrap,
parametric bootstrap, out-of-bag bootstrap, and repeated 10-, 5-, and
2-fold cross-validation for Elastic Net logistic regression and Random
Forest.

The remainder of this paper is organised as follows.
Section~\ref{sec-methods} describes the simulation design,
model-development procedures, resampling procedures, and performance
measures. Section~\ref{sec-results} presents the results, and
Section~\ref{sec-discussion} discusses the findings and their
implications.

\section{Methods}\label{sec-methods}

We present the design of our simulation study following the ADEMP
framework (Morris et al. 2019), in terms of the Aims, Data-Generating
Mechanism, Estimands and other targets, Methods, and Performance
Measures.

\subsection{Aim}\label{aim}

The aim of the simulation study is to evaluate how well six practical
resampling methods estimate the stability of individual predictions when
combined with regression-based and tree-based model-development
procedures. We compared the non-parametric bootstrap, parametric
bootstrap, out-of-bag bootstrap, and repeated 10-, 5-, and 2-fold
cross-validation against repeated independent sampling of the same
sample size, which serves as the simulation reference for
individual-level prediction instability. Of course, in practice when
developing a model, repeated independent sampling is not possible, but
the premise of bootstrapping or cross-validation is to closely replicate
that process.

\subsection{Data-Generating Mechanism}\label{data-generating-mechanism}

We considered two data-generating mechanisms in our simulation: one
based on a linear model, and a second based on a Random Forest model.
Predictor variables were generated in the same fashion for both the
linear and non-linear data-generating mechanisms.

Let \(\mathbf{X}\) denote a vector of \(K=12\) continuous predictors
generated from a multivariate normal distribution:

\[
\mathbf{X} \sim MVN(\boldsymbol{\mu}, \boldsymbol{\Sigma}),
\]

with mean \(\boldsymbol{\mu}=\mathbf{0}\) and covariance
\(\boldsymbol{\Sigma}\), which has unit variances and pairwise
correlations of \(\rho=0.2\) for all predictors.

\subsubsection{Linear Data-Generating
Mechanism}\label{linear-data-generating-mechanism}

In our linear data-generating mechanism we defined the linear predictor
as:

\[
\eta
= w_0
+ w_1 \sum_{k=1}^{6} x_k
+ 3w_1 \sum_{k=7}^{9} x_k,
\]

with intercept \(w_0 = -1.4649\) and covariate weights \(w_1 = 0.1610\).
Half of the predictor variables had (relatively weak) effects on the
outcome equal to \(w_1\), one fourth had relatively strong effects
(three times as large as the weak effects), and one fourth had no effect
on the outcome. The intercept and predictor effects (\(w_0\) and
\(w_1\)) were optimised such that they yielded a desired c-statistic
(AUC) of 0.80 (see Section~\ref{sec-regcoef-optimization} for details).

The true probability of the outcome occurring was generated using a
logistic transformation:

\begin{equation}\protect\phantomsection\label{eq-logtrans}{
p = \frac{e^\eta}{1 + e^\eta}.
}\end{equation}

Finally, the binary outcome \(y\) was generated from a Bernoulli
distribution with outcome probability \(p\):

\begin{equation}\protect\phantomsection\label{eq-bernoulli}{
y \sim \operatorname{Bernoulli}(p),
}\end{equation}

where 0 and 1 indicate absence and occurrence of the outcome,
respectively.

\subsubsection{Non-Linear Data-Generating
Mechanism}\label{non-linear-data-generating-mechanism}

In our non-linear data-generating mechanism, a Random Forest model was
used as the data-generating model. This data-generating Random Forest
model was developed using a preparatory training dataset with sample
size \(n = 1000\). This preparatory training dataset was generated with
defined non-linearities using the following linear predictor:

\begin{equation}\protect\phantomsection\label{eq-eta-nl}{
\eta = w_0 + w_1 \left( x_1*x_2 - x_3*x_4 + x_5^4 + x_6^4 \right) + 3w_1 \left( \, x_7*x_8 - x_9^4 \right).
}\end{equation}

The intercept \(w_0\) and covariate weights \(w_1\) were set again to
-1.4649 and 0.1610, respectively. Similar to the linear data-generating
mechanism, half of the predictor variables had (relatively weak) effects
on the outcome equal to \(w_1\), one fourth had relatively strong
effects (three times as large as the weak effects), and one fourth had
no effect on the outcome. Finally, according to
Equation~\ref{eq-logtrans} and Equation~\ref{eq-bernoulli}, a true
outcome probability and binary outcome were generated for the
preparatory training dataset.

Using this (non-linear) training dataset, the data-generating Random
Forest model was trained to predict a binary outcome according to the
procedure detailed in Section~\ref{sec-methods} (all Random Forest
models in our simulation were trained according to the same procedure).

The fitted Random Forest model was then used as the non-linear DGM
throughout the simulation study: each individual's true outcome
probability was generated using the data-generating Random Forest model

\[
p = f_{\mathrm{RF}}(\mathbf{x}),
\] and the binary outcome \(y\) was generated from a Bernoulli
distribution with probability \(p\).

\subsubsection{Simulation Scenarios}\label{simulation-scenarios}

Under each data-generating mechanism we considered two sample-size
scenarios, \(N\in\{229, 458\}\) where the larger sample size corresponds
to the minimum required sample size as calculated using the approach
proposed by Riley et al. (2019) and implemented in the R package
\textbf{pmsampsize} (Ensor (2023)), and the smaller sample size is half
that. The target outcome fraction was set to \(0.25\), resulting in an
expected 9.5 (4.8) events per predictor parameter under the larger
(smaller) sample size. The four simulation scenarios are summarised in
Table~\ref{tbl-sims}.

\begin{table}[H]

\caption{\label{tbl-sims}Overview of Simulation Scenarios.}

\centering{

\centering
\begin{threeparttable}
\begin{tabular}[t]{lcccc}
\toprule
 & Scenario 1 & Scenario 2 & Scenario 3 & Scenario 4\\
\midrule
Data-Generating Mechanism & Linear & Linear & Non-linear & Non-linear\\
Sample Size & 458 & 229 & 458 & 229\\
Target Outcome Fraction\textsuperscript{*} & 0.25 & 0.25 & 0.275 & 0.275\\
Number of Predictor Parameters & 12 & 12 & 12 & 12\\
\bottomrule
\end{tabular}
\begin{tablenotes}[para]
\item[*] For the non-linear DGM, the outcome fraction refers to the logistic model used to generate the Random Forest's initial training data.
\end{tablenotes}
\end{threeparttable}

}

\end{table}%

\subsection{Estimands}\label{estimands}

For each individual in an initial dataset, the reference target was the
distribution of outcome probability estimates obtained by applying the
complete model-development procedure to independent samples of size
\(N\) from the data-generating mechanism (DGM). This distribution
includes variation from both model fitting and hyperparameter tuning. We
assessed how closely each resampling procedure recovered selected
characteristics of this distribution, including its variance, empirical
95\% range, and mean absolute differences from the original prediction
and the true outcome probability (DGM probability). The DGM probability
\(p\) was additionally used as the ground truth for assessing prediction
accuracy.

\subsection{Methods}\label{methods}

This study followed a factorial simulation design with two sample sizes,
two model-development procedures, two DGMs, and six practical resampling
procedures: non-parametric bootstrap, parametric bootstrap, out-of-bag
bootstrap, and repeated 10-, 5-, and 2-fold cross-validation. The
model-development procedures were Elastic Net logistic regression and
Random Forest. A summary of the simulation scenarios is provided in
Table~\ref{tbl-sims}.

\subsubsection{Simulation Procedure}\label{simulation-procedure}

For each sample-size setting and DGM, 100 initial datasets were
generated. Elastic Net and Random Forest models were fitted to each
initial dataset and used to obtain an in-sample original prediction for
every individual. Using each of the six resampling procedures as well as
independent sampling, resampled versions of each of the 100 initial
datasets were created (see Table~\ref{tbl-resampling-methods}), and the
model-development procedure was repeated in each resample to obtain a
prediction for every individual in the corresponding initial dataset.
The resampling procedures were implemented to produce 200 predictions
per individual, model, and initial dataset.

\subsubsection{Prediction Models}\label{prediction-models}

We evaluated two binary-outcome prediction models: Elastic Net logistic
regression and Random Forest. Elastic Net combines L1 (Lasso) and L2
(Ridge) penalties, where \(\lambda_{EN}\) controls the penalty strength
and \(\alpha_{EN}\) controls the balance between the two penalties (Zou
and Hastie 2005). Random Forest combines probability trees grown from
bootstrap samples and random subsets of candidate predictors. The
implemented Random Forest used 500 trees and tuned the number of
candidate predictors at each split (\(\mathrm{mtry}\)) and the minimum
terminal-node size (\(\mathrm{min.node.size}\)).

For the linear data-generating mechanism, Random Forest and Elastic Net
models were fitted using all 12 candidate predictors. For the non-linear
data-generating mechanism, all 12 candidate predictors were used to fit
the Random Forest models, while for the Elastic Net models, candidate
predictors were first transformed such that the model equation aligned
with the non-linear data-generating mechanism presented in
Equation~\ref{eq-eta-nl}. This ensured that both models were similarly
capable of capturing the non-linearities in the data. Although such
non-linearities are rarely known in advance in practice, our aim is to
assess how well resampling methods estimate individual-level prediction
stability. We therefore used models with comparable ability to fit the
data so that differences in model misspecification did not confound the
comparison. We compared the fit of the initial models prior to beginning
the resampling procedures, as detailed in
Section~\ref{sec-performance-measures}.

\subsubsection{Hyperparameter Tuning}\label{sec-hyperparameter-tuning}

The choice of hyperparameters to tune and their search ranges was
informed by Probst et al. (2019). Hyperparameter tuning was repeated
every time a model was fitted. Thus, the simulation evaluated the
stability of the complete tuning-and-fitting procedure rather than
stability conditional on the hyperparameters selected in the initial
dataset.

For Elastic Net, \(\alpha_{EN}\) was evaluated on the grid
\(0.0,0.1,\ldots,1.0\). A single random 10-fold split was reused across
these values. For each value of \(\alpha_{EN}\), the value of
\(\lambda_{EN}\) minimising the binomial deviance was identified. The
\(\alpha_{EN}\) with the lowest recorded cross-validated deviance and
its corresponding selected \(\lambda_{EN}\) were used to fit the final
model. The final model was fitted to all available training
observations.

For Random Forest, 30 candidate pairs were randomly sampled with
replacement, with \(\mathrm{mtry}\in\{1,\ldots,12\}\) and
\(\mathrm{min.node.size}\in\{1,\ldots,20\}\). Candidates were evaluated
using 10-fold cross-validated log loss, with 500 probability trees
fitted for each candidate-fold combination. The pair with the lowest
mean log loss was used to fit a final 500-tree probability forest to all
available training observations.

\subsubsection{Resampling Strategies}\label{resampling-strategies}

The implementation of each resampling procedure and the
independent-sampling reference is described in
Table~\ref{tbl-resampling-methods}.

\begin{longtable}[t]{>{\raggedright\arraybackslash}p{4.2cm}>{\raggedright\arraybackslash}p{10.0cm}}

\caption{\label{tbl-resampling-methods}Overview of methods for
estimating stability of individual predictions.}

\tabularnewline

\toprule
Method & Description\\
\midrule
\endfirsthead
\multicolumn{2}{@{}l}{\textit{(continued)}}\\
\toprule
Method & Description\\
\midrule
\endhead

\endfoot
\bottomrule
\endlastfoot
Independent sampling & Generate 200 independent datasets of size $N$ from the true data-generating mechanism. Fit models to each and predict for individuals in the initial dataset. This serves as the baseline for ideal stability estimation.\\
\addlinespace
Non-parametric bootstrap & Draw 200 bootstrap samples of size $N$ with replacement from the initial dataset. Fit models to each resample and predict for all individuals in the initial dataset.\\
\addlinespace
Parametric bootstrap & Fit each model-development procedure to the initial dataset and obtain model-specific fitted probabilities. For each of 200 replicates, simulate a separate Bernoulli outcome vector from each model's fitted probabilities while keeping the predictor matrix fixed. Refit each model to its own simulated outcomes and predict all individuals in the initial dataset.\\
\addlinespace
Out-of-bag bootstrap & Draw bootstrap samples of size $N$ with replacement, fit both models, and predict only for the individuals absent from each sample. Continue until each individual has 200 out-of-bag predictions.\\
\addlinespace
10-fold cross-validation & Repeat 10-fold cross-validation 200 times. In each repetition, fit 10 models, each using nine folds to predict the remaining fold, so each individual receives one prediction from a model fitted to approximately $0.9N$ observations in each of the 200 repetitions.\\
\addlinespace
5-fold cross-validation & Analogous to repeated 10-fold cross-validation, using approximately $0.8N$ observations for each fit.\\
\addlinespace
2-fold cross-validation & Analogous to repeated 10-fold cross-validation, using approximately $0.5N$ observations for each fit.\\*

\end{longtable}

\subsection{Performance Measures}\label{sec-performance-measures}

We first evaluated the initial models and datasets. To quantify
in-sample discriminative ability, we calculated the c-statistic (AUC)
for both models in each initial dataset, and summarised it within each
sample-size and DGM scenario. In-sample ``calibration'' was assessed per
model, sample-size and DGM scenario by plotting the true (DGM)
probability against the estimated probability and fitting a LOESS
smoother, pooling all observations across individuals and initial
datasets. For out-of-sample performance, we calculated AUCs for the
models fitted to each independent sample and applied to the
corresponding initial dataset. The AUC distribution per initial dataset
was visualised using boxplots and summarised, pooled across initial
datasets, by its mean and 2.5\% and 97.5\% percentiles. Out-of-sample
calibration curves were constructed using the models fitted in one
independent sample per initial dataset and visualised using initial
dataset-specific and pooled LOESS calibration curves. Corresponding
figures were created for each of the resampling procedures.

Using the 200 predictions generated from independent sampling as well as
each of the six resampling procedures for each individual, model, and
initial dataset, we calculated the empirical variance; the width of the
empirical 95\% range, defined as the 97.5th minus the 2.5th percentile;
the mean absolute difference from the original in-sample prediction
(MAPE); and the mean absolute difference from the true (DGM) probability
(true MAPE; tMAPE).

To characterise the independent-sampling reference, instability metrics
were summarised per DGM, sample size, model, and metric by averaging
across individuals and initial datasets. To visualise heterogeneity in
individual-level instability over the true (DGM) probability range, we
fitted generalised additive model smooths separately per DGM, sample
size, model and metric.

Finally, we visually assessed recovery of the variance, 95\% range
width, MAPE, and tMAPE by comparing each resampling procedure with
independent sampling for the same individual, model, and initial
dataset. For the variance, 2.5th and 97.5th percentiles of the
prediction distribution, 95\% range width, MAPE, and tMAPE, we
calculated the signed difference (resampling procedure minus independent
sampling) and visualised them across the true (DGM) outcome probability
range. For variance, 95\% range width, MAPE, and tMAPE, we calculated
the individual-level absolute relative deviation as
\(|\mathrm{estimate} - \mathrm{reference}| / \mathrm{reference}\), using
the corresponding independent-sampling value as the reference. We fitted
generalised additive model smooths to these absolute relative deviations
across the true (DGM) outcome probability range, pooling individuals and
initial datasets separately by DGM, sample size, model, metric, and
resampling procedure. These curves estimate the conditional mean
absolute relative deviation at a given true outcome probability.

\subsection{Software}\label{software}

The simulation was implemented in R version 4.6.0 and run using
high-performance computing infrastructure at UMC Utrecht. Elastic Net
models were fitted using the \texttt{glmnet} package (version 5.0,
Friedman et al. (2010), Tay et al. (2023)), and Random Forest models
were fitted using the \texttt{ranger} package (version 0.18.0, Wright
and Ziegler (2017)).

\FloatBarrier

\section{Results}\label{sec-results}

\subsection{Simulation completeness and initial model
performance}\label{simulation-completeness-and-initial-model-performance}

For each sample-size and DGM scenario, 100 initial datasets were
generated. The realised mean outcome fractions were approximately 25\%
under the linear DGM (for both sample sizes) and 28\% and 29\% under the
non-linear DGM for the larger and smaller sample size, respectively
(S-Table \ref{tbl-app-initial-performance}).

The initially fitted Elastic Net models had mean in-sample AUCs of 0.81
for N = 458 and 0.82 for N = 229 under the linear DGM, and mean AUCs of
0.77 under the non-linear DGM (for either sample size). In line with
expectations, the in-sample Random Forest AUCs were overly optimistic:
1.00 in all sample-size and DGM scenarios. When models fitted to
independent samples were applied to the corresponding initial datasets,
mean AUCs in N = 458 were 0.79 for Elastic Net and 0.77 for Random
Forest under the linear DGM. The corresponding mean AUCs under the
non-linear DGM were 0.74 and 0.71, respectively. For both DGMs, the
corresponding AUCs in N = 229 were slightly lower (S-Table
\ref{tbl-app-initial-performance}). Distributions of these AUCs across
independently fitted models are shown in
S-Figure~\ref{sfig-independent-sample-auc}.

Under the linear DGM, average DGM probabilities closely followed the
initially fitted Elastic Net predictions, with some overprediction at
higher estimated probabilities for the smaller sample size. Under the
non-linear DGM, agreement was less uniform, with underprediction at
intermediate estimated probabilities and overprediction at the upper
end. Random Forest predictions showed substantially greater dispersion
relative to DGM probabilities under both mechanisms and generally
underestimated outcome probabilities at low predicted values and
overestimated them at higher values (S-Figure~\ref{sfig-calibration};
S-Figure~\ref{sfig-calibration-nonlin}). Outcome-based calibration of
models fitted to independent samples was closer to the diagonal for the
linear DGM, although some overprediction remained at higher estimated
probabilities (S-Figure~\ref{sfig-calibration-indep}). Under the
non-linear DGM, calibration departures were more pronounced for Elastic
Net, whereas Random Forest showed smaller departures over much of the
estimated probability range
(S-Figure~\ref{sfig-calibration-indep-nonlin}).

Calibration plots for models fitted using the bootstrap and
cross-validation procedures under both DGMs are provided in S-Figures
\ref{sfig-calibration-boot-lin}--\ref{sfig-calibration-cv-nonlin}. The
distributions of hyperparameters selected for the models fitted to the
initial datasets are shown in Section~\ref{sec-hyperpars}.

\subsection{Prediction instability under independent
sampling}\label{prediction-instability-under-independent-sampling}

Under independent sampling, individual-level prediction variance,
empirical 95\% range width, MAPE and tMAPE mean absolute prediction
error relative to the original estimate and true outcome probability,
respectively) were, on average, larger for Random Forest than for
Elastic Net (Table~\ref{tbl-results-independent-reference}). Smaller
development samples and the non-linear DGM were also associated with
higher average values across these measures. The increase from the
linear to the non-linear DGM was substantially greater for tMAPE than
for MAPE, indicating that distance from true outcome probability and
distance from the original probability estimate changed differently.

\begin{figure}

\centering{

\pandocbounded{\includegraphics[keepaspectratio]{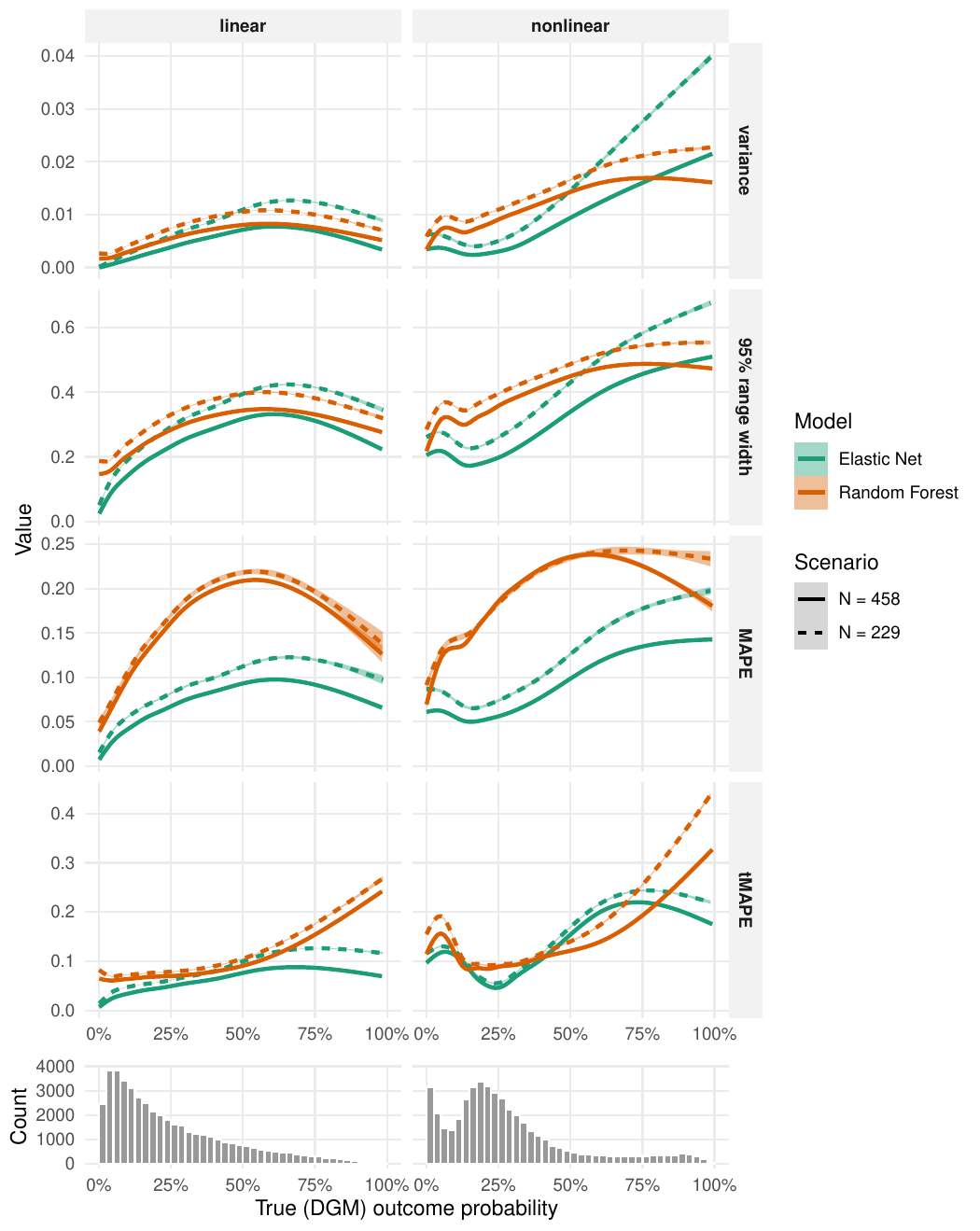}}

}

\caption{\label{fig-results-independent-heterogeneity}Individual-level
prediction instability under independent sampling across the true (DGM)
outcome probability range. Lines are generalised additive model smooths
fitted separately by model and scenario; shaded bands show 95\%
confidence intervals. Histograms below the panels show the distribution
of true probabilities across individuals and initial datasets for the
larger sample size.}

\end{figure}

Figure~\ref{fig-results-independent-heterogeneity} shows smoothed
individual-level prediction-instability measures across true (DGM)
outcome probability, pooled across individuals and initial datasets.
Under the linear DGM, variance, empirical 95\% range width, and MAPE
generally increased towards intermediate-to-high true probabilities and
then declined. Under the non-linear DGM, variance and range width
continued to increase at higher true probabilities for Elastic Net,
whereas they generally levelled off or declined for Random Forest. MAPE
followed a similar pattern, increasing towards the upper end of the
probability range for Elastic Net but peaking at intermediate-to-high
probabilities for Random Forest. Under both DGMs, tMAPE increased
sharply at high true probabilities for Random Forest, whereas it peaked
at intermediate-to-high probabilities for Elastic Net. Smaller
development samples generally yielded higher values, although the Random
Forest MAPE curves were similar across sample sizes over much of the
probability range.

\begin{table}

\caption{\label{tbl-results-independent-reference}Prediction instability
under independent sampling. Values are means {[}empirical 2.5th, 97.5th
percentiles{]} of the individual-averaged measure across 100 initial
datasets.}

\centering{

\centering
\begin{tabular}[t]{lccl}
\toprule
DGM & metric & N = 458 & N = 229\\
\midrule
\addlinespace[0.3em]
\multicolumn{4}{l}{\textbf{Elastic Net}}\\
\hspace{1em} & variance & 0.0034 [0.0030, 0.0036] & 0.0055 [0.0049, 0.0063]\\
\arrayrulecolor{lightgray}\cmidrule{2-4}\arrayrulecolor{black}

\hspace{1em} & 95\% range width & 0.201 [0.189, 0.210] & 0.260 [0.245, 0.278]\\
\arrayrulecolor{lightgray}\cmidrule{2-4}\arrayrulecolor{black}

\hspace{1em} & MAPE & 0.059 [0.048, 0.074] & 0.074 [0.060, 0.094]\\
\arrayrulecolor{lightgray}\cmidrule{2-4}\arrayrulecolor{black}

\hspace{1em}\multirow{-4}{*}{\raggedright\arraybackslash linear} & tMAPE & 0.047 [0.044, 0.049] & 0.063 [0.059, 0.069]\\
\arrayrulecolor{gray}\cmidrule{1-4}\arrayrulecolor{black}

\hspace{1em} & variance & 0.0053 [0.0046, 0.0060] & 0.0089 [0.0076, 0.0103]\\
\arrayrulecolor{lightgray}\cmidrule{2-4}\arrayrulecolor{black}

\hspace{1em} & 95\% range width & 0.243 [0.229, 0.257] & 0.315 [0.298, 0.336]\\
\arrayrulecolor{lightgray}\cmidrule{2-4}\arrayrulecolor{black}

\hspace{1em} & MAPE & 0.070 [0.056, 0.094] & 0.092 [0.073, 0.131]\\
\arrayrulecolor{lightgray}\cmidrule{2-4}\arrayrulecolor{black}

\hspace{1em}\multirow{-4}{*}{\raggedright\arraybackslash nonlinear} & tMAPE & 0.098 [0.093, 0.104] & 0.111 [0.102, 0.119]\\
\cmidrule{1-4}

\addlinespace[0.3em]
\multicolumn{4}{l}{\textbf{Random Forest}}\\
\hspace{1em} & variance & 0.0047 [0.0044, 0.0051] & 0.0064 [0.0058, 0.0071]\\
\arrayrulecolor{lightgray}\cmidrule{2-4}\arrayrulecolor{black}

\hspace{1em} & 95\% range width & 0.253 [0.243, 0.264] & 0.297 [0.282, 0.311]\\
\arrayrulecolor{lightgray}\cmidrule{2-4}\arrayrulecolor{black}

\hspace{1em} & MAPE & 0.136 [0.101, 0.204] & 0.146 [0.107, 0.210]\\
\arrayrulecolor{lightgray}\cmidrule{2-4}\arrayrulecolor{black}

\hspace{1em}\multirow{-4}{*}{\raggedright\arraybackslash linear} & tMAPE & 0.077 [0.073, 0.080] & 0.087 [0.082, 0.093]\\
\arrayrulecolor{gray}\cmidrule{1-4}\arrayrulecolor{black}

\hspace{1em} & variance & 0.0096 [0.0090, 0.0101] & 0.0120 [0.0113, 0.0128]\\
\arrayrulecolor{lightgray}\cmidrule{2-4}\arrayrulecolor{black}

\hspace{1em} & 95\% range width & 0.361 [0.351, 0.371] & 0.406 [0.394, 0.420]\\
\arrayrulecolor{lightgray}\cmidrule{2-4}\arrayrulecolor{black}

\hspace{1em} & MAPE & 0.174 [0.139, 0.227] & 0.179 [0.133, 0.240]\\
\arrayrulecolor{lightgray}\cmidrule{2-4}\arrayrulecolor{black}

\hspace{1em}\multirow{-4}{*}{\raggedright\arraybackslash nonlinear} & tMAPE & 0.115 [0.110, 0.121] & 0.136 [0.127, 0.144]\\
\bottomrule
\end{tabular}

}

\end{table}%

\subsection{Recovery of individual-level prediction-instability
metrics}\label{recovery-of-individual-level-prediction-instability-metrics}

Having characterised prediction instability under independent sampling,
we next assessed how closely each resampling procedure recovered the
individual-level metrics.

\begin{figure}

\centering{

\pandocbounded{\includegraphics[keepaspectratio]{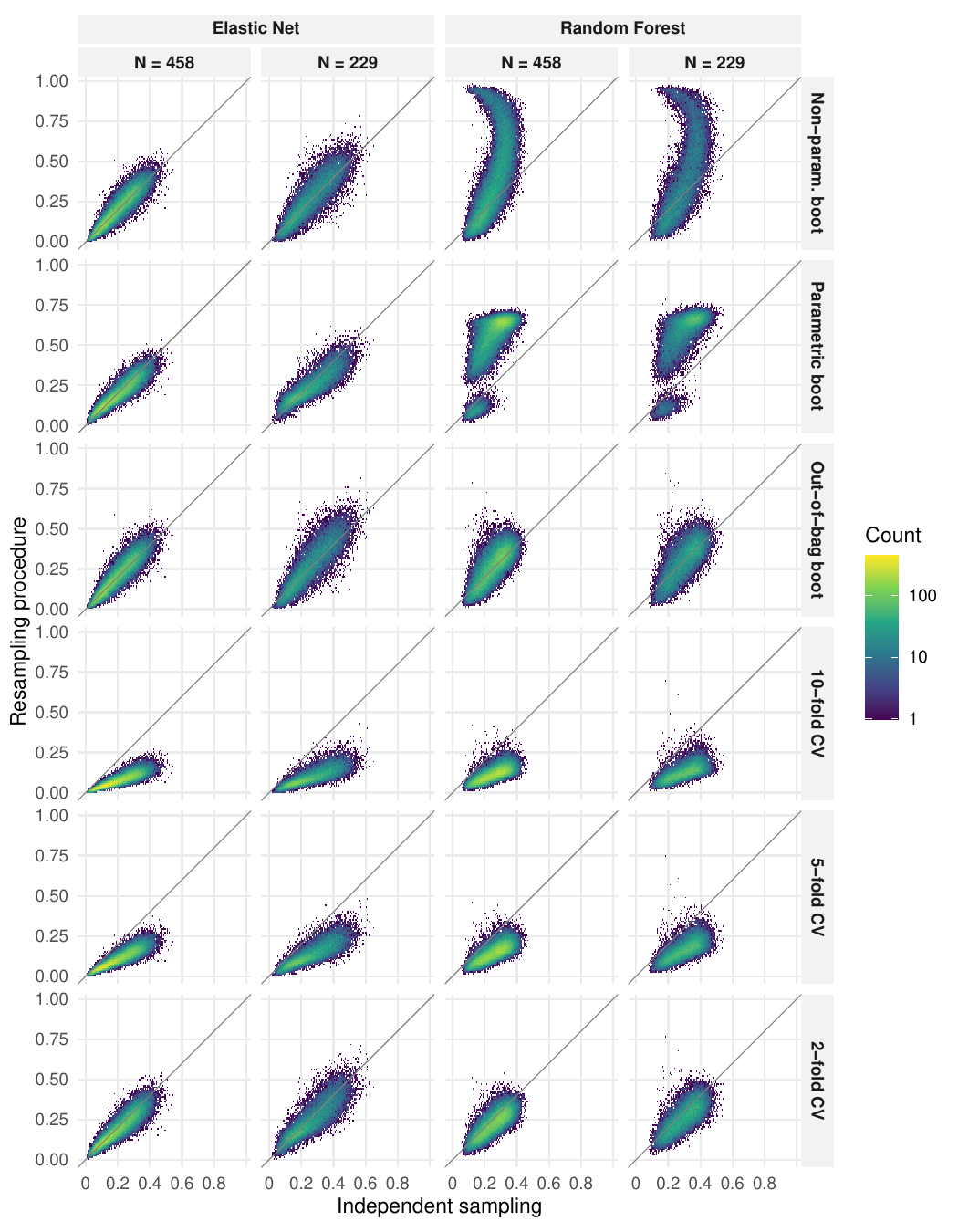}}

}

\caption{\label{fig-results-agreement-width}Individual-level agreement
with independent sampling for empirical 95\% range width under the
linear DGM.}

\end{figure}

\begin{figure}

\centering{

\pandocbounded{\includegraphics[keepaspectratio]{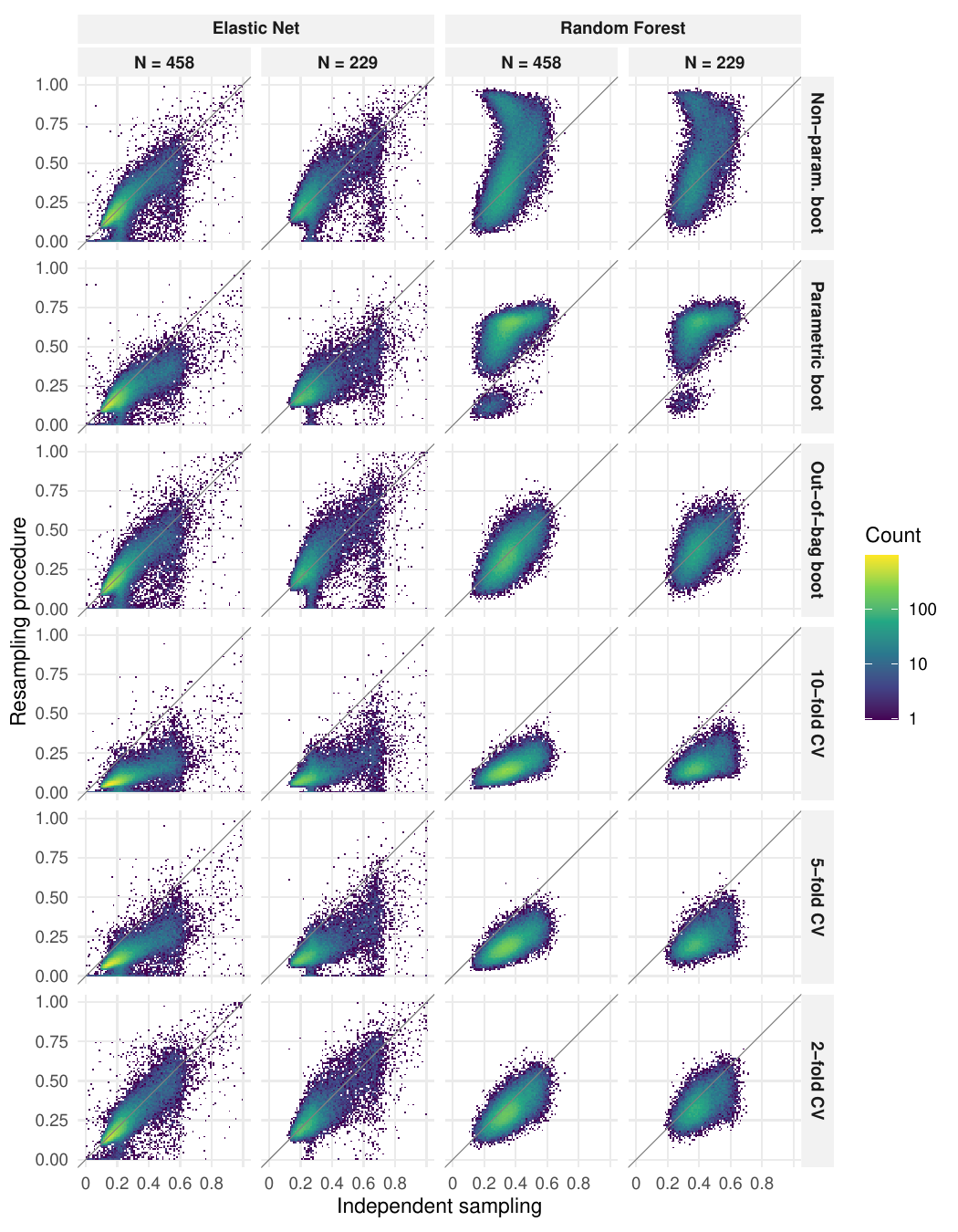}}

}

\caption{\label{fig-results-agreement-width-nonlin}Individual-level
agreement with independent sampling for empirical 95\% range width under
the non-linear DGM.}

\end{figure}

Across the agreement plots for empirical range width
(Figure~\ref{fig-results-agreement-width};
Figure~\ref{fig-results-agreement-width-nonlin}) and the supplementary
plots for MAPE, variance, and tMAPE (S-Figures \ref{sfig-agreement-mape}
- \ref{sfig-agreement-tmape-nonlin}), recovery depended on both the
resampling procedure and model. For Elastic Net, the bootstrap
procedures and repeated 2-fold cross-validation were generally closest
to the diagonal, whereas repeated 5- and 10-fold cross-validation tended
to underestimate empirical range width. For Random Forest, repeated
2-fold cross-validation and the out-of-bag bootstrap showed the closest
agreement for empirical range width. The non-parametric and especially
the parametric bootstrap tended to overestimate empirical range width,
whereas repeated 5- and 10-fold cross-validation tended to underestimate
it. Under the non-linear DGM, parametric bootstrap also tended to
underestimate Elastic Net range width. For Random Forest, out-of-bag
bootstrap and repeated 2-fold cross-validation remained closer to the
reference, although both tended to underestimate range width under this
DGM. Prediction variance showed similar patterns to empirical 95\% range
width (S-Figure~\ref{sfig-agreement-variance} and
S-Figure~\ref{sfig-agreement-variance-nonlin}).

MAPE and tMAPE showed different patterns of agreement, particularly for
Random Forest. Both bootstrap procedures often underestimated MAPE
relative to independent sampling, especially for individuals with large
reference MAPE, but overestimated tMAPE (S-Figures
\ref{sfig-agreement-mape}--\ref{sfig-agreement-mape-nonlin} and
\ref{sfig-agreement-tmape}--\ref{sfig-agreement-tmape-nonlin}). Thus,
bootstrap predictions could be closer to the original in-sample
prediction while being farther from the true DGM probability.

\begin{figure}

\centering{

\pandocbounded{\includegraphics[keepaspectratio]{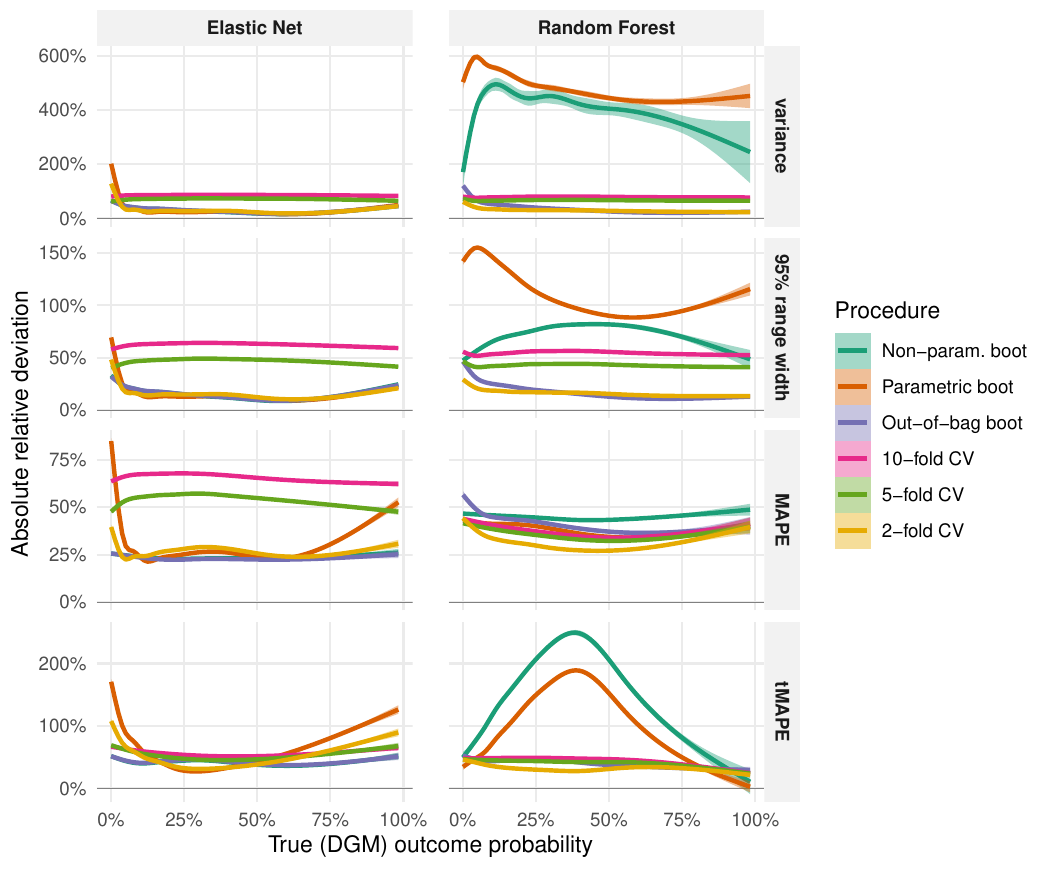}}

}

\caption{\label{fig-results-risk-patterns}Absolute relative
individual-level deviation from independent sampling across the true
(DGM) outcome probability range for the linear DGM in the larger
sample-size scenario. Each individual-level absolute difference was
divided by the corresponding independent-sampling reference value. Lines
are generalised additive model estimates of the conditional mean, fitted
separately by resampling procedure and model; shaded bands show 95\%
confidence intervals.}

\end{figure}

\begin{figure}

\centering{

\pandocbounded{\includegraphics[keepaspectratio]{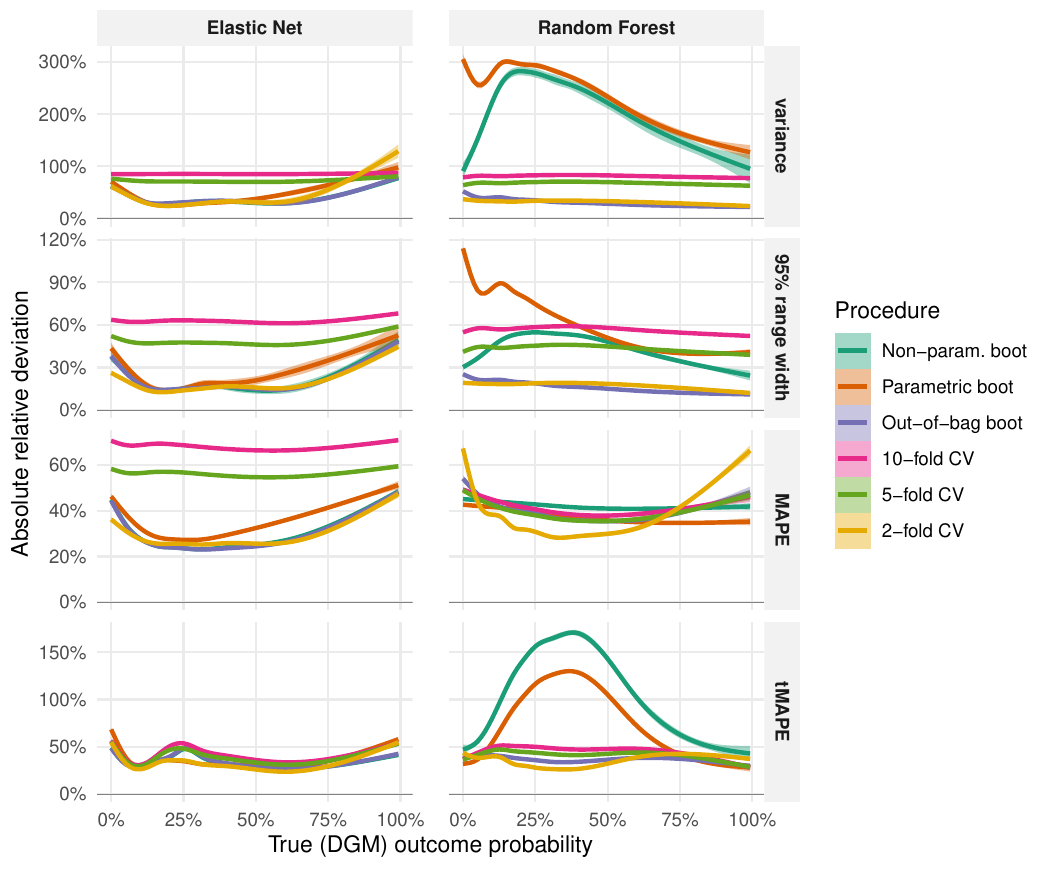}}

}

\caption{\label{fig-results-risk-patterns-nonlin}Absolute relative
individual-level deviation from independent sampling across the true
(DGM) outcome probability range for the non-linear DGM in the larger
sample-size scenario. Each individual-level absolute difference was
divided by the corresponding independent-sampling reference value. Lines
are generalised additive model estimates of the conditional mean, fitted
separately by resampling procedure and model; shaded bands show 95\%
confidence intervals. One outlying observation was excluded (see
Section~\ref{sec-risk-pattern-outlier}).}

\end{figure}

Absolute relative deviations varied by procedure, metric, and true
outcome probability (Figure~\ref{fig-results-risk-patterns};
Figure~\ref{fig-results-risk-patterns-nonlin};
S-Figure~\ref{sfig-risk-patterns-smaller};
S-Figure~\ref{sfig-risk-patterns-smaller-nonlin}). For default
non-parametric bootstrap, deviations were generally smaller for Elastic
Net than for Random Forest. For Elastic Net, the default non-parametric
bootstrap was fairly stable in the linear DGM and showed somewhat
increasing deviation for larger probabilities, and was in both DGMs and
for all four instability metrics very similar to out-of-bag bootstrap.
For Random Forest models, however, deviations of default non-parametric
bootstrap based variance, 95\% range width and tMAPE were much larger
than those of out-of-bag bootstrap. These deviations were more
pronounced for the larger sample size and were highest around 11\%
outcome probability for the variance (deviation approx. 495\%), around
45\% for the 95\% range width (deviation approx. 82\%) and around 38\%
for the tMAPE (deviation approx. 249\%). Deviations for MAPE in Random
Forest models were relatively constant across probabilities and more
similar to those of out-of-bag bootstrap.

Parametric bootstrap generally showed larger Random Forest deviations
than non-parametric bootstrap for variance and range width, but smaller
deviations for MAPE and tMAPE. Its tMAPE curve also peaked at
intermediate probabilities, reaching a maximum of 189\% at 39\% outcome
probability for the larger sample size and the linear DGM. For Elastic
Net under the linear DGM, parametric-bootstrap tMAPE deviations
increased towards both ends of the probability range.

The out-of-bag bootstrap closely resembled default non-parametric
bootstrap for Elastic Net, but showed substantially smaller Random
Forest variance and range-width deviations and no pronounced
intermediate-probability tMAPE peak. Repeated 2-fold cross-validation
also showed comparatively small variance and range-width deviations
across much of the probability range but had higher deviations under the
linear DGM for Elastic Net at low true probabilities. Under the
non-linear DGM, Elastic Net variance and 95\% range deviations increased
at higher probabilities, particularly in the larger sample, as did
Random Forest MAPE deviations.

Repeated 5- and 10-fold cross-validation showed persistent, relatively
flat variance and range-width deviations across models, DGMs, and sample
sizes, with larger deviations for 10-fold cross-validation (variance:
75\%--87\%; 95\% range width: 51\%--68\%) than for 5-fold
cross-validation (variance: 62\%--81\%, 95\% range-width: 39\%--59\%).
Their MAPE deviations were also pronounced for Elastic Net, but 5- and
10-fold cross-validation had similar MAPE deviations for Random Forest
and similar tMAPE deviations for both models. Smaller sample size did
not consistently increase relative deviations.

Results for the 2.5th and 97.5th percentiles (S-Figures
\ref{sfig-agreement-q025} - \ref{sfig-agreement-q975-nonlin}) showed
whether procedures recovered both the location and width of the
prediction distribution. Repeated 5- and 10-fold cross-validation tended
to overestimate the 2.5th percentile and underestimate the 97.5th
percentile, consistent with distributions that were too narrow. For
Random Forest, parametric bootstrap generally underestimated the 2.5th
percentile and overestimated the 97.5th percentile. Non-parametric
bootstrap also tended to underestimate the lower endpoint, but
upper-endpoint errors were more heterogeneous, with pronounced
overestimation for a subset of individuals.

Supplementary signed-difference plots
(S-Figure~\ref{sfig-original-risk-patterns-glmnet};
S-Figure~\ref{sfig-original-risk-patterns-glmnet-nonlin};
S-Figure~\ref{sfig-original-risk-patterns-rf};
S-Figure~\ref{sfig-original-risk-patterns-rf-nonlin}) show non-linear,
sometimes sign-changing errors, especially for the Random Forest
bootstrap procedures.

\subsection{Illustrative individual-level prediction
distributions}\label{sec-illustrative-prediction-distributions}

\begin{landscape}

\begin{figure}

\centering{

\pandocbounded{\includegraphics[keepaspectratio]{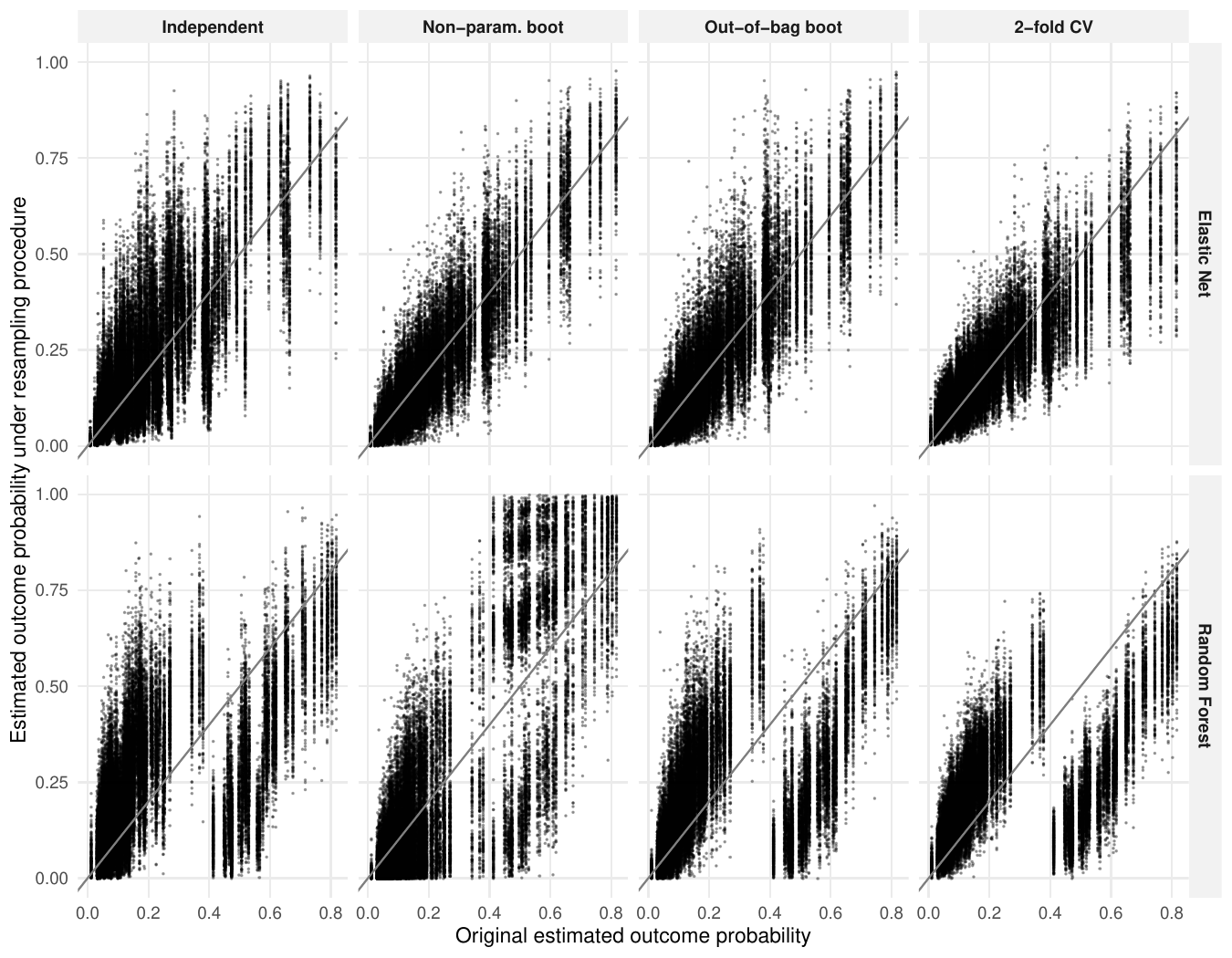}}

}

\caption{\label{fig-prediction-distributions}Illustrative prediction
instability plot showing individual-level prediction distributions for
selected procedures under the linear DGM in one initial dataset of the
smaller sample-size scenario. Each individual contributes 200
predictions per procedure. The diagonal line denotes equality with the
original estimated outcome probability.}

\end{figure}%

\end{landscape}

\begin{landscape}

\begin{figure}

\centering{

\pandocbounded{\includegraphics[keepaspectratio]{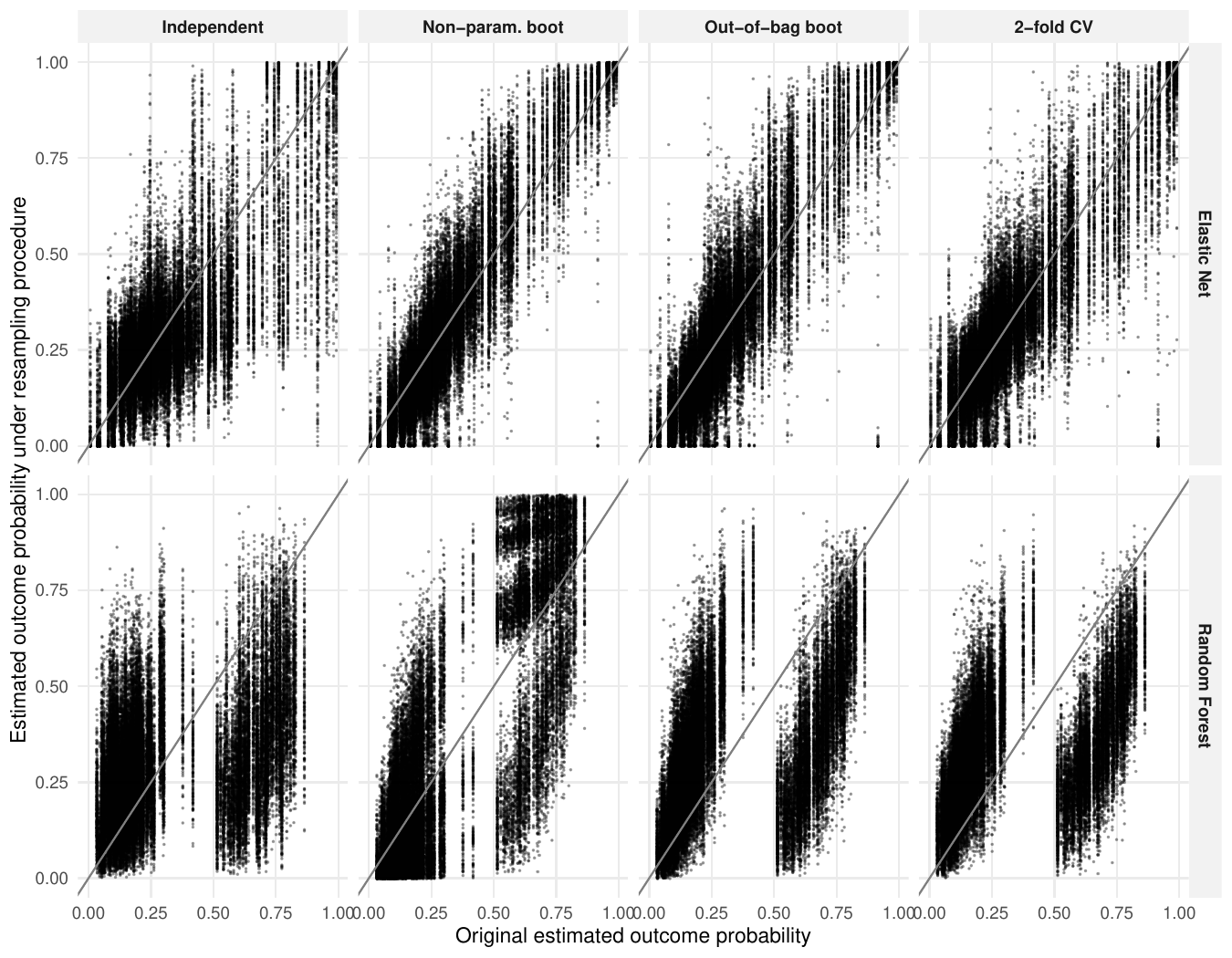}}

}

\caption{\label{fig-prediction-distributions-nonlin}Illustrative
prediction instability plot showing individual-level prediction
distributions for selected procedures under the non-linear DGM in one
initial dataset of the smaller sample-size scenario. Each individual
contributes 200 predictions per procedure. The diagonal line denotes
equality with the original estimated outcome probability.}

\end{figure}%

\end{landscape}

For one initial dataset of 229 individuals under each DGM,
Figure~\ref{fig-prediction-distributions} and
Figure~\ref{fig-prediction-distributions-nonlin} show prediction
instability plots of the 200 predictions per individual obtained using
independent sampling, default non-parametric bootstrap, out-of-bag
bootstrap, and repeated 2-fold cross-validation. The vertical spread at
each individual's original estimated outcome probability illustrates
variation across fitted models. Differences in both the spread and
location of the predictions were visible between independent sampling
and the resampling procedures for both models. These differences were
particularly pronounced for default non-parametric bootstrap with Random
Forest, which produced many predictions concentrated near zero or one.

\FloatBarrier

\section{Discussion}\label{sec-discussion}

\subsubsection*{Principal findings}\label{principal-findings}
\addcontentsline{toc}{subsubsection}{Principal findings}

In this simulation study, we compared how well six practical resampling
procedures recovered selected characteristics of the individual-specific
prediction instability obtained under repeated independent sampling.
None performed uniformly well across the two model-development
procedures, the distributional characteristics, and the full range of
individual probabilities. For Elastic Net, bootstrap procedures and
repeated 2-fold cross-validation generally recovered prediction
variance, empirical range width and MAPE more closely than repeated 5-
or 10-fold cross-validation, although relative deviations varied with
DGM and true outcome probability. For Random Forest, default
non-parametric and parametric bootstrap often overestimated variance,
range width and tMAPE, but underestimated MAPE. Repeated 5- and 10-fold
cross-validation generally underestimated prediction spread. Repeated
2-fold cross-validation and out-of-bag bootstrap were often closer to
the reference, although neither was uniformly best across measures and
outcome probabilities.

Taken together, these findings show that resampling procedures are not
interchangeable estimators of prediction instability. Importantly, less
variable predictions are not necessarily more accurate; they may instead
underestimate the variability that would arise if model development were
repeated in new samples.

Agreement with independent sampling also varied across the true outcome
probability range. Consequently, average measures of agreement across
individuals can conceal substantial underestimation or overestimation of
instability for particular outcome probability profiles.

The larger difference between Random Forest and Elastic Net for MAPE
than for tMAPE under independent sampling reflects the different
reference probabilities used by these measures. MAPE measures distance
from the original in-sample prediction, whereas tMAPE measures distance
from the true DGM probability. For Random Forest, the original
predictions were drawn towards the observed outcomes and often differed
substantially from the underlying probabilities. Predictions from
independently fitted models could therefore be closer to the true
outcome probability while remaining farther from the original
prediction, consistent with the lower average tMAPE than MAPE. The
comparatively good independent-sampling calibration over much of the
probability range also illustrates why adaptation to training outcomes
should be distinguished from poor generalisation, as discussed by Curth
(2024).

Thus, the larger MAPE partly reflects its outcome-influenced reference
prediction, rather than solely greater spread in predictions across
fitted models. Our Random Forest results apply to the tuned procedure
examined here and should not be generalised to all tree-based or
machine-learning algorithms. They are consistent with previous
observations that Random Forest probability estimates can be sensitive
to local data characteristics and model complexity (Barreñada et al.
2024).

\subsubsection*{Why the resampling procedures
differ}\label{why-the-resampling-procedures-differ}
\addcontentsline{toc}{subsubsection}{Why the resampling procedures
differ}

For Elastic Net, the default non-parametric bootstrap performed
comparatively well, which is consistent with its intended role as an
empirical approximation to repeated sampling. To represent the complete
data-adaptive model-development procedure, we reselected both the
penalty type and strength within every resample. This is an essential
feature of the assessment: fixing the tuning choices selected in the
initial dataset would condition on those data-driven choices and omit
their contribution to prediction instability. Uncertainty in tuning
parameters, particularly in smaller samples, is therefore part of the
instability that the procedure should capture (Riley and Collins 2023;
Riley et al. 2021). While our results support the default non-parametric
bootstrap as a useful approximation for this Elastic Net procedure under
the conditions examined, its poorer agreement with independent sampling
for Random Forest shows that this performance does not extend uniformly
across model-development procedures.

The default bootstrap mixes predictions from models fitted with and
without the target individual, whereas independently sampled models
never include that individual. This distinction may be particularly
relevant for Random Forest, whose in-sample predictions were drawn
towards the observed outcomes. Differences between predictions for
included and excluded individuals may contribute to the wider bootstrap
prediction distributions and larger tMAPE observed relative to
independent sampling. We did not isolate this mechanism directly, so it
remains an interpretation rather than a demonstrated causal explanation
that warrants investigation in future research.

The out-of-bag bootstrap excludes the target individual from model
fitting, bringing it closer to independent sampling and potentially
explaining its better agreement for several instability measures.
However, it still samples from the original empirical distribution and
includes repeated observations and fewer than \(N\) unique individuals.
Its relatively good performance may therefore reflect compensating
differences rather than faithful reproduction of independent sampling.

Repeated cross-validation predicts each individual while held out, but
repeatedly partitions the same finite dataset into folds. The resulting
training sets overlap substantially, especially with more folds, so
prediction variability is expected to be lower than under independent
sampling. Cross-validation also reduces training-sample size to
approximately \(0.9N\), \(0.8N\), and \(0.5N\) for 10-, 5-, and 2-fold
cross-validation, respectively. Fewer folds therefore increase expected
instability through both smaller training sets and reduced overlap.
Repeated 5- and 10-fold cross-validation underestimated variability in
our simulations, whereas 2-fold cross-validation agreed more closely
with independent sampling. This closer agreement may reflect
compensation between smaller training sets and reuse of the original
data; it does not imply that 2-fold cross-validation targets the
independent sampling distribution by construction.

Our findings contextualise the empirical comparison by Khongthon et al.
(2026), which found lower MAPE with repeated 5-fold cross-validation
than with bootstrap resampling, particularly for Random Forest. Their
study demonstrated differences between instability estimates in clinical
data; our independent-sampling reference shows why lower estimated
instability need not indicate better approximation. Repeated 5-fold
cross-validation often underestimated prediction variability in our
simulations. Together, these findings emphasise that comparisons and
selection of resampling procedures should be guided by the distribution
they are intended to approximate.

Unlike independent sampling, the parametric bootstrap fixes the observed
predictors and generates outcomes from the original fitted probabilities
rather than the true DGM. It therefore retains distortions in those
probabilities, particularly relevant for Random Forest, whose original
predictions were drawn towards observed outcomes. Our design cannot
separate the contributions of fixed predictors, fitted outcome
probabilities, and variability during refitting to its poor agreement
with independent sampling. Generating new outcomes alone therefore does
not ensure a closer approximation.

By contrast, Thomassen et al. (2025) found that the parametric bootstrap
more accurately estimated effective sample size, which is inversely
related to prediction variance at a given true outcome probability.
Their non-parametric bootstrap performed well for Elastic Net but
substantially overestimated effective sample size for Random Forest,
implying underestimated prediction variance. In our study, bootstrap
methods generally performed well for Elastic Net, but both often
overestimated Random Forest variance, particularly the parametric
bootstrap. These findings concern different sampling experiments:
Thomassen et al. (2025) used a fitted-model DGM with fixed predictors
and hyperparameters, whereas we generated independent datasets from the
true DGM, varying both predictors and outcomes and repeating tuning.

Computational cost also differed between procedures. In our
implementation, repeated 2-fold cross-validation took approximately
twice as long as the default non-parametric bootstrap, while repeated
5-fold cross-validation and the parametric bootstrap took approximately
three times as long. Repeated 10-fold cross-validation and the
out-of-bag bootstrap required approximately four to six times as long.
Thus, the closer agreement obtained with some procedures came at greater
computational cost. These approximate comparisons depend on the
implementation and computing environment and should not be interpreted
as general benchmarks.

\subsubsection*{Strengths and
Limitations}\label{strengths-and-limitations}
\addcontentsline{toc}{subsubsection}{Strengths and Limitations}

A strength of our simulation design is the ability to compare resampling
estimates with corresponding estimates obtained from genuinely
independent development samples. Such a reference is unavailable in
empirical applications and allowed us to distinguish the magnitude of
prediction instability from how accurately each procedure estimated it.
Complementary instability measures, linear and non-linear DGMs with
appropriately specified candidate predictors, and repeated
hyperparameter tuning broadened the comparison and captured variation in
the complete development procedure.

However, the simulation covered only two model types, two
data-generating mechanisms and two sample sizes. Although instability is
known to decrease with larger development samples (Riley and Collins
2023; Riley et al. 2025), it remains to be established whether the
relative performance of the resampling procedures holds in much smaller
and much larger samples. Results may differ with categorical or
high-dimensional predictors, greater class imbalance, missing data,
other types of outcomes, tuning strategies, or different degrees of
learner regularisation. Because the in-sample Random Forest predictions
were drawn towards the observed outcomes, comparisons with more strongly
regularised forests or other flexible learners would be particularly
informative. Development pipelines that include explicit feature
selection, for example followed by unpenalised logistic regression, are
also common in clinical prediction modelling, and assessing their
instability would be particularly relevant for practice.

Additional work is required to fully isolate the mechanism by which the
resampling procedures differ from independent sampling. For
cross-validation, we could not separate the effects of smaller training
samples from those of observation reuse. For the default bootstrap,
comparing predictions from models that included versus excluded the
target individual would help assess whether target-individual inclusion
explains the Random Forest results. The parametric bootstrap
simultaneously fixes predictors and simulates model-specific outcomes
from fitted probabilities, preventing attribution of discrepancies to
any one component.

Each empirical 95\% range was estimated from 200 predictions, leaving
few observations in each tail. Although 200 resamples has been proposed
as a practical minimum (Riley and Collins 2023), a sensitivity analysis
with more replicates would help assess Monte Carlo error in
percentile-based measures. The consistent patterns for range width and
variance provide some reassurance about the qualitative findings.

Finally, we assessed sampling-related uncertainty within a fixed
population and a prespecified development procedure. We did not quantify
uncertainty arising from population or measurement changes, alternative
modelling choices, or lack of applicability to a new setting (Riley et
al. 2025). Estimating prediction instability does not remove these
sources of uncertainty or establish that a model is sufficiently
accurate or useful for clinical decision making.

\subsubsection*{Conclusion}\label{conclusion}
\addcontentsline{toc}{subsubsection}{Conclusion}

Practical resampling procedures can give materially different
representations of individual-level prediction instability. While in our
simulation default bootstrap recovered prediction instability relatively
well for Elastic Net, it often overestimated prediction variability in
Random Forest models, in some cases, dramatically. Such model-dependent
estimation errors may distort comparisons of prediction instability
between model-development procedures, even when the same resampling
method is applied to each. Therefore, non-parametric bootstrapping
should not be used as a one-size-fits-all method for estimating
individual prediction instability. In our simulation, we found that
methods that did not use any training data in their predictions while
limiting overlap between repeated resamples (i.e., out-of-bag bootstrap
and repeated 2-fold cross-validation) produced estimates of prediction
instability that were more often closer to the reference for both
models. Further research may investigate if these methods can be used to
provide reasonable approximations of prediction instability for a wider
range of prediction modelling strategies and under different data
scenarios.

\section*{Funding}\label{funding}
\addcontentsline{toc}{section}{Funding}

This publication was made possible through funding from the Sector Plan
for Medicine and Health Sciences, supported by the Dutch Ministry of
Education, Culture and Science (OCW). We also gratefully acknowledge
funding by ZonMW VIDI project number 09150172310078.

\newpage{}

\appendix

\setcounter{table}{0}
\renewcommand{\tablename}{S-Table}

\FloatBarrier

\section{Supplementary material}\label{sec-supp-results}

\subsection{Regression coefficient
optimisation}\label{sec-regcoef-optimization}

The intercept and regression coefficients were optimised numerically to
achieve a target outcome fraction and an approximate target AUC of 0.25
and 0.80, respectively. We generated predictor structures for 20
datasets with \(N=100{,}000\) observations per dataset. Exploiting the
common coefficient structure of the data-generating mechanisms, the
predictor score of individual \(i\) could be expressed as
\[\eta_{i} = w_0 + w_1 z_{i},\] where \(z_{i}\) is the linear
combination of covariate terms using relative coefficient multipliers of
1 for weak terms and 3 for strong terms; null predictors were omitted.
For the linear data-generating mechanism, these terms were the covariate
values themselves, whereas for the non-linear mechanism they included
transformations and interactions.

The calibration of \(w_0\) and \(w_1\) to the target AUC and outcome
rate was performed using a nested procedure.

The outer optimisation searched over candidate values of \(w_1\) using
the \texttt{optim} function in R with the L-BFGS-B algorithm, an initial
value of 0.5 and lower and upper bounds of 0.01 and 10, respectively.
For each candidate value of \(w_1\), we determined the value of \(w_0\)
that brought the average outcome probability across all observations in
the 20 datasets closest to the target outcome fraction using
\texttt{uniroot}:

\[
\frac{1}{20N}\sum_{j=1}^{20}\sum_{i=1}^{N}
\frac{\exp\left(w_0+w_1z_{ij}\right)}{1+\exp\left(w_0+w_1z_{ij}\right)}=0.25.
\]

For each candidate value of \(w_1\) and its corresponding intercept
\(w_0\), we calculated the outcome probabilities for each dataset and
generated binary outcomes from Bernoulli distributions with these
probabilities.

We then estimated the AUC separately in each dataset using the generated
outcomes and outcome probabilities, and calculated the squared
difference between the achieved AUC and the target AUC. The squared
differences were averaged and used as the loss value for that candidate:

\[
L(w_1)=\frac{1}{20}\sum_{j=1}^{20}
\left[\widehat{\operatorname{AUC}}_j(w_1)-0.80\right]^2.
\]

The loss value was subsequently used by L-BFGS-B to guide the iterative
search for \(w_1\).

Once the final value of \(w_1\) was returned by the optimiser, the
intercept \(w_0\) was recalculated to obtain the final calibrated
intercept--coefficient pair.

\newpage{}

\subsection{Additional tables and figures}\label{sec-additional-results}

The additional tables and figures are organised as follows.

\subsubsection*{Model performance and
calibration}\label{model-performance-and-calibration}
\addcontentsline{toc}{subsubsection}{Model performance and calibration}

\begin{itemize}
\tightlist
\item
  S-Table \ref{tbl-app-initial-performance} summarises outcome fractions
  and in-sample and independent-sample AUCs;
\item
  S-Figure~\ref{sfig-independent-sample-auc} shows the distributions of
  independent-sample AUCs.
\item
  Agreement between the initially fitted predictions and the DGM
  probabilities is shown in S-Figure~\ref{sfig-calibration} (linear DGM)
  and S-Figure~\ref{sfig-calibration-nonlin} (non-linear DGM).
\item
  Calibration plots for models fitted to independent samples, bootstrap
  samples and cross-validation training samples are shown in S-Figures
  \ref{sfig-calibration-indep}--\ref{sfig-calibration-indep-nonlin},
  S-Figures
  \ref{sfig-calibration-boot-lin}--\ref{sfig-calibration-boot-nonlin},
  and S-Figures
  \ref{sfig-calibration-cv-lin}--\ref{sfig-calibration-cv-nonlin},
  respectively.
\end{itemize}

\subsubsection*{Individual-level agreement with independent
sampling}\label{individual-level-agreement-with-independent-sampling}
\addcontentsline{toc}{subsubsection}{Individual-level agreement with
independent sampling}

\begin{itemize}
\tightlist
\item
  S-Figure~\ref{sfig-risk-patterns-smaller} shows how absolute relative
  deviations in instability metrics vary across the DGM probability
  range in the smaller sample-size scenario.
\end{itemize}

The remaining agreement plots show

\begin{itemize}
\tightlist
\item
  MAPE (S-Figures
  \ref{sfig-agreement-mape}--\ref{sfig-agreement-mape-nonlin}),
\item
  prediction variance (S-Figures
  \ref{sfig-agreement-variance}--\ref{sfig-agreement-variance-nonlin}),
\item
  tMAPE (S-Figures
  \ref{sfig-agreement-tmape}--\ref{sfig-agreement-tmape-nonlin}),
\item
  the lower and upper endpoints of the empirical 95\% prediction range
  (S-Figures
  \ref{sfig-agreement-q025}--\ref{sfig-agreement-q975-nonlin}).
\end{itemize}

\subsubsection*{Patterns across the true DGM probability
range}\label{patterns-across-the-true-dgm-probability-range}
\addcontentsline{toc}{subsubsection}{Patterns across the true DGM
probability range}

Differences from independent sampling in the larger sample-size scenario
are shown for Elastic Net (S-Figures
\ref{sfig-original-risk-patterns-glmnet}--\ref{sfig-original-risk-patterns-glmnet-nonlin})
and Random Forest (S-Figures
\ref{sfig-original-risk-patterns-rf}--\ref{sfig-original-risk-patterns-rf-nonlin}).

\FloatBarrier

\newpage{}

\begin{landscape}

\begin{table}

\caption{\label{tbl-app-initial-performance}Model performance in the
original datasets. In-sample AUCs are summarised across 100 initial
datasets. Independent-sample AUCs are based on 200 independently fitted
models per initial dataset, each applied to the corresponding original
dataset. Values are means {[}empirical 2.5th, 97.5th percentiles{]}.}

\centering{

\centering
\begin{tabular}[t]{lllccc}
\toprule
\makecell[c]{Sample\\size} & DGM & Outcome fraction & Model & \makecell[c]{AUC\\in-sample} & \makecell[c]{AUC\\independent-sample}\\
\midrule
 &  &  & Elastic Net & 0.810 [0.774, 0.852] & 0.787 [0.747, 0.829]\\
\arrayrulecolor{lightgray}\cmidrule{4-6}\arrayrulecolor{black}

 & \multirow{-2}{*}{\raggedright\arraybackslash linear} & \multirow{-2}{*}{\raggedright\arraybackslash 0.250 [0.213, 0.289]} & Random Forest & 0.998 [0.992, 1.000] & 0.767 [0.719, 0.813]\\
\arrayrulecolor{gray}\cmidrule{2-6}\arrayrulecolor{black}

 &  &  & Elastic Net & 0.765 [0.716, 0.814] & 0.737 [0.668, 0.794]\\
\arrayrulecolor{lightgray}\cmidrule{4-6}\arrayrulecolor{black}

\multirow{-4}{*}[0.5\dimexpr\aboverulesep+\belowrulesep+\cmidrulewidth]{\raggedright\arraybackslash N = 458} & \multirow{-2}{*}{\raggedright\arraybackslash nonlinear} & \multirow{-2}{*}{\raggedright\arraybackslash 0.283 [0.242, 0.325]} & Random Forest & 0.999 [0.993, 1.000] & 0.714 [0.651, 0.772]\\

 &  &  & Elastic Net & 0.820 [0.758, 0.889] & 0.779 [0.710, 0.851]\\
\arrayrulecolor{lightgray}\cmidrule{4-6}\arrayrulecolor{black}

 & \multirow{-2}{*}{\raggedright\arraybackslash linear} & \multirow{-2}{*}{\raggedright\arraybackslash 0.249 [0.197, 0.302]} & Random Forest & 0.998 [0.991, 1.000] & 0.758 [0.680, 0.832]\\
\arrayrulecolor{gray}\cmidrule{2-6}\arrayrulecolor{black}

 &  &  & Elastic Net & 0.768 [0.665, 0.834] & 0.713 [0.607, 0.794]\\
\arrayrulecolor{lightgray}\cmidrule{4-6}\arrayrulecolor{black}

\multirow{-4}{*}[0.5\dimexpr\aboverulesep+\belowrulesep+\cmidrulewidth]{\raggedright\arraybackslash N = 229} & \multirow{-2}{*}{\raggedright\arraybackslash nonlinear} & \multirow{-2}{*}{\raggedright\arraybackslash 0.289 [0.225, 0.356]} & Random Forest & 0.998 [0.991, 1.000] & 0.678 [0.590, 0.762]\\
\bottomrule
\end{tabular}

}

\end{table}%

\end{landscape}

\begin{suppfig}

\centering{

\pandocbounded{\includegraphics[keepaspectratio]{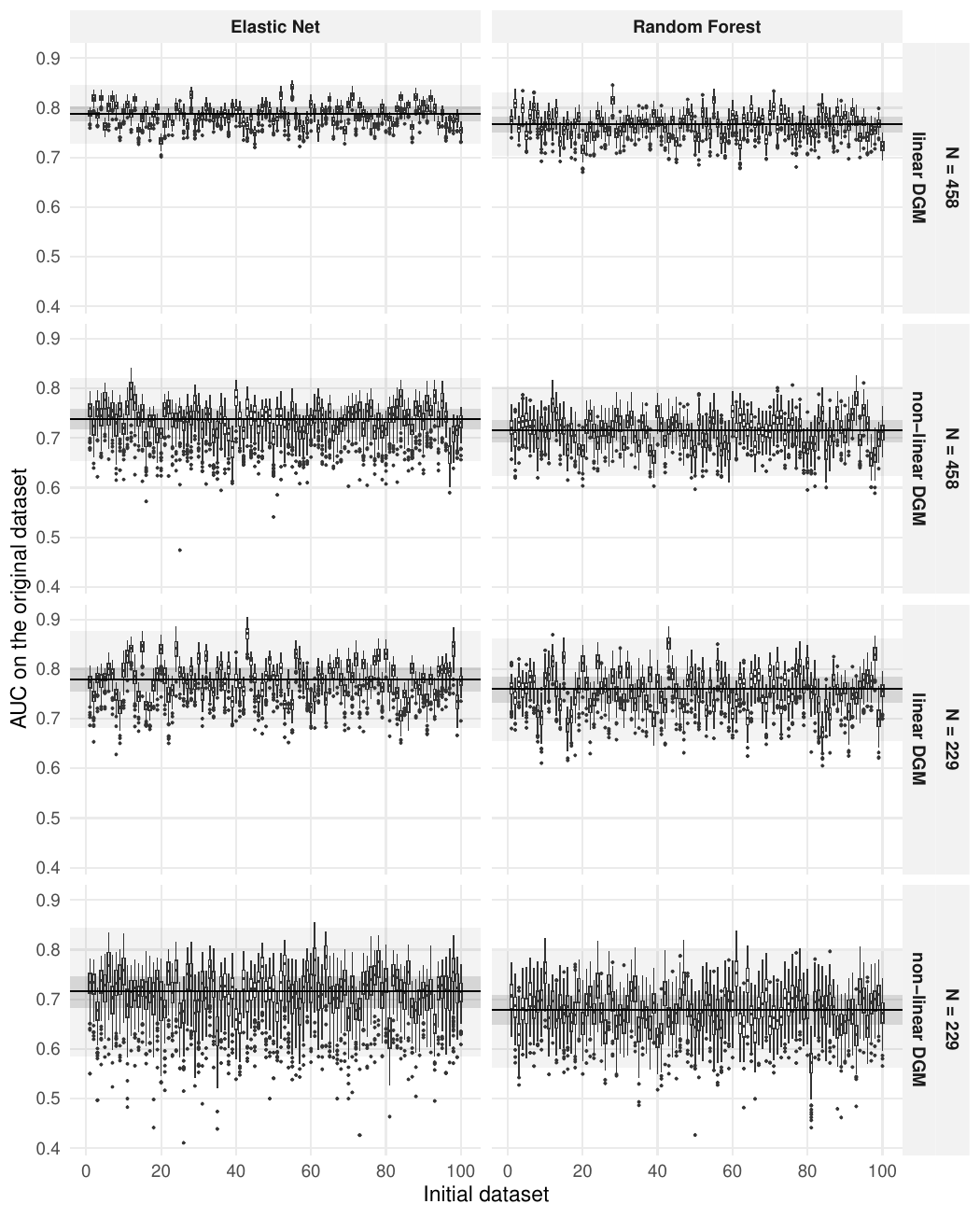}}

}

\caption{\label{sfig-independent-sample-auc}AUCs of models fitted to
independently sampled development data and applied to the initial
datasets. The boxplots show the 200 AUC values per initial dataset.
Within each panel, the horizontal line and grey bands show the pooled
median, interquartile range and standard (1.5 IQR) boxplot-whisker
span.}

\end{suppfig}%

\begin{suppfig}

\centering{

\includegraphics[width=0.9\linewidth,height=\textheight,keepaspectratio]{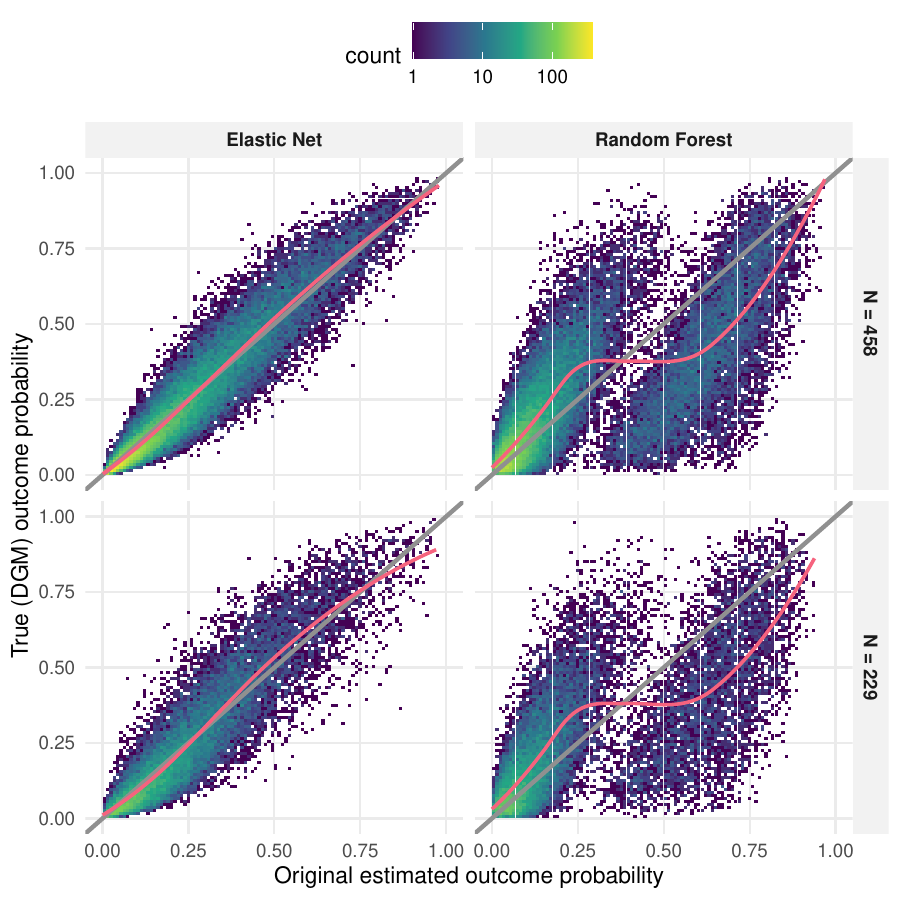}

}

\caption{\label{sfig-calibration}Agreement between in-sample predictions
and DGM probabilities across initially fitted models for the linear DGM.
The red lines show a LOESS curve.}

\end{suppfig}%

\begin{suppfig}

\centering{

\includegraphics[width=0.9\linewidth,height=\textheight,keepaspectratio]{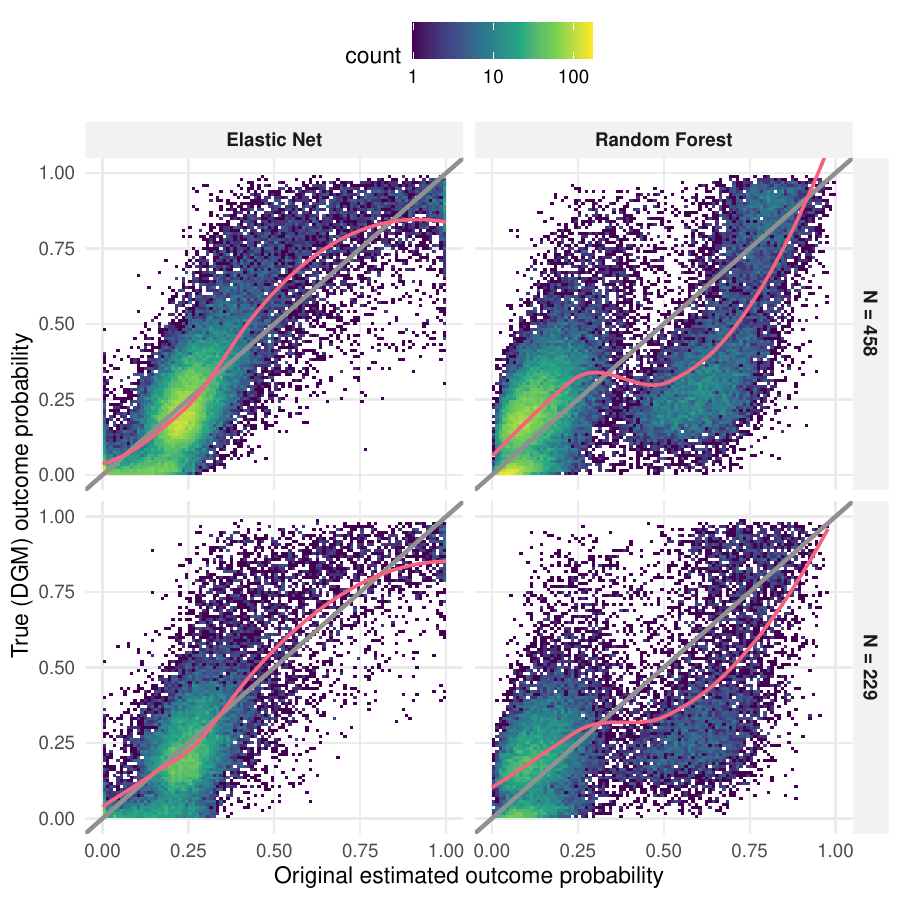}

}

\caption{\label{sfig-calibration-nonlin}Agreement between in-sample
predictions and DGM probabilities across initially fitted models for the
non-linear DGM. The red lines show a LOESS curve.}

\end{suppfig}%

\begin{suppfig}

\centering{

\includegraphics[width=0.8\linewidth,height=\textheight,keepaspectratio]{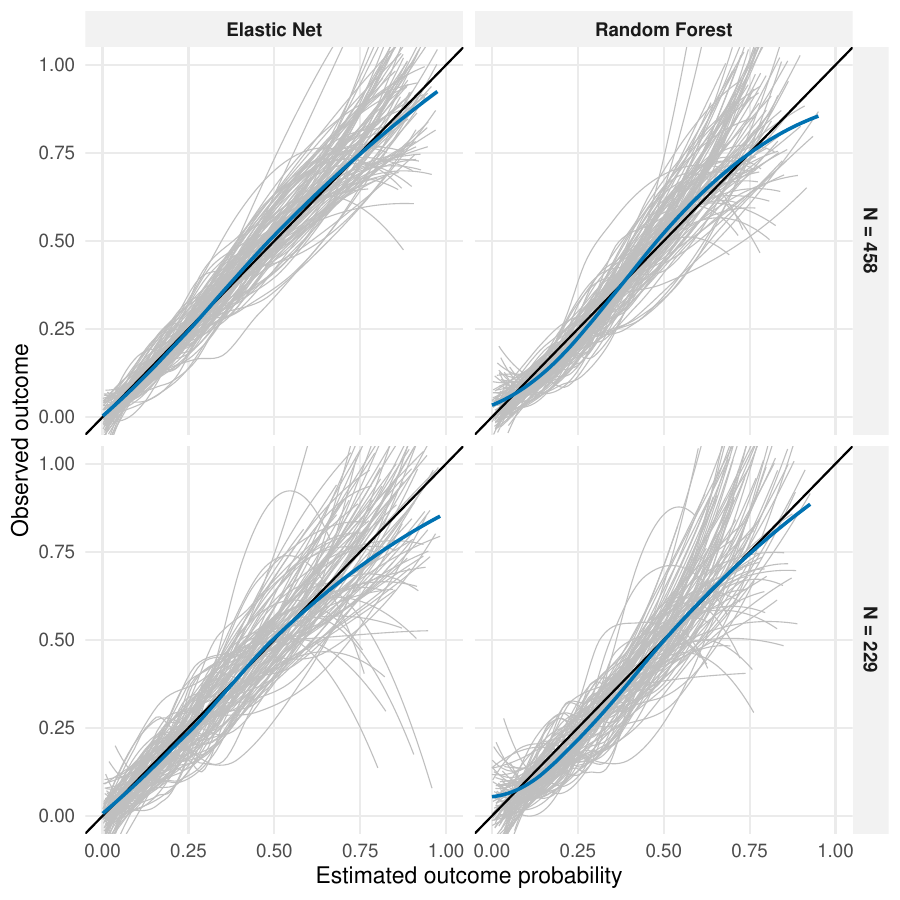}

}

\caption{\label{sfig-calibration-indep}Calibration of the models fitted
on the first independent sample for the linear DGM. Thin grey lines show
initial dataset-specific LOESS curves; blue lines show pooled LOESS
curves within each scenario and model.}

\end{suppfig}%

\begin{suppfig}

\centering{

\includegraphics[width=0.8\linewidth,height=\textheight,keepaspectratio]{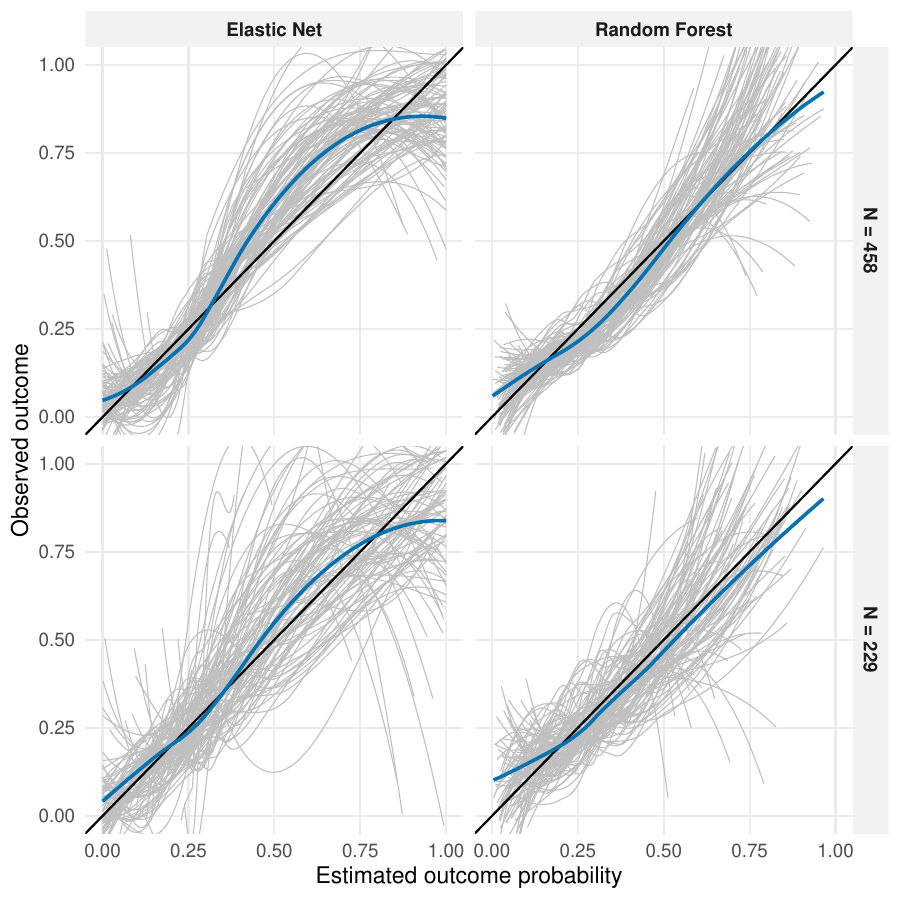}

}

\caption{\label{sfig-calibration-indep-nonlin}Calibration of the models
fitted on the first independent sample for the non-linear DGM. Thin grey
lines show initial dataset-specific LOESS curves; blue lines show pooled
LOESS curves within each scenario and model.}

\end{suppfig}%

\begin{suppfig}

\centering{

\includegraphics[width=0.8\linewidth,height=\textheight,keepaspectratio]{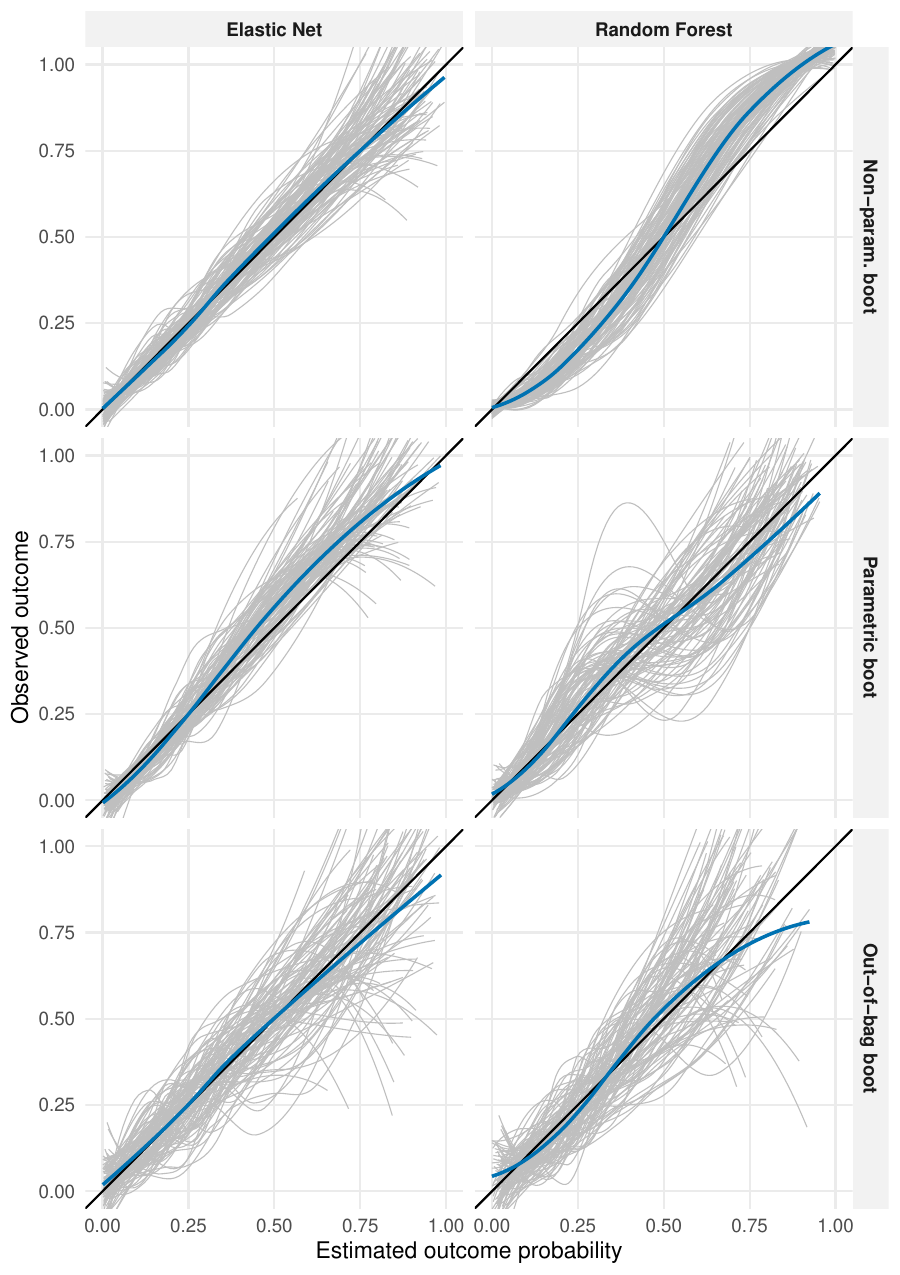}

}

\caption{\label{sfig-calibration-boot-lin}Calibration of the models
fitted on the first non-parametric, parametric and out-of-bag bootstrap
samples for the linear DGM. Thin grey lines show initial
dataset-specific LOESS curves; blue lines show pooled LOESS curves
within each scenario and model.}

\end{suppfig}%

\begin{suppfig}

\centering{

\includegraphics[width=0.8\linewidth,height=\textheight,keepaspectratio]{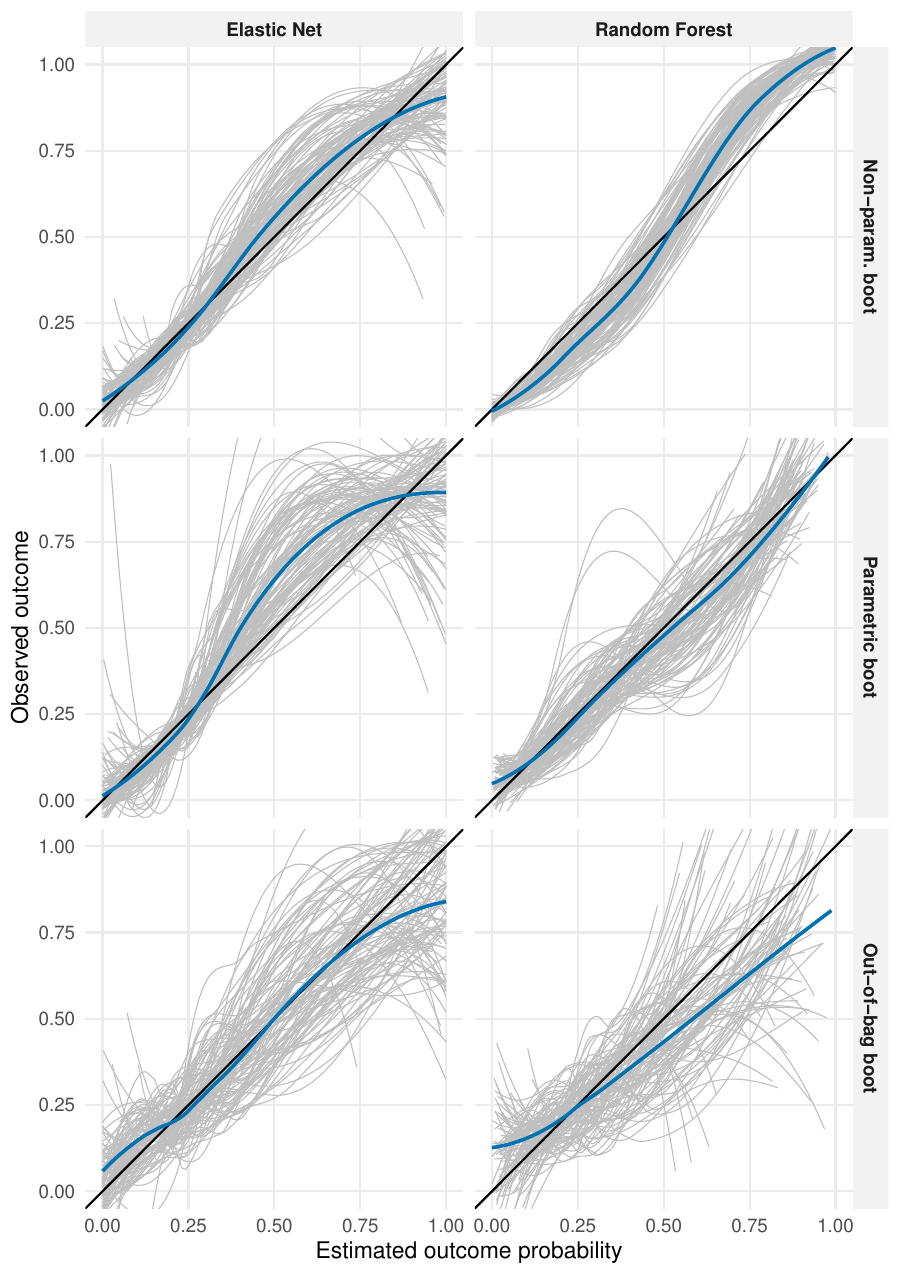}

}

\caption{\label{sfig-calibration-boot-nonlin}Calibration of the models
fitted on the first non-parametric, parametric and out-of-bag bootstrap
samples for the non-linear DGM. Thin grey lines show initial
dataset-specific LOESS curves; blue lines show pooled LOESS curves
within each scenario and model.}

\end{suppfig}%

\begin{suppfig}

\centering{

\includegraphics[width=0.8\linewidth,height=\textheight,keepaspectratio]{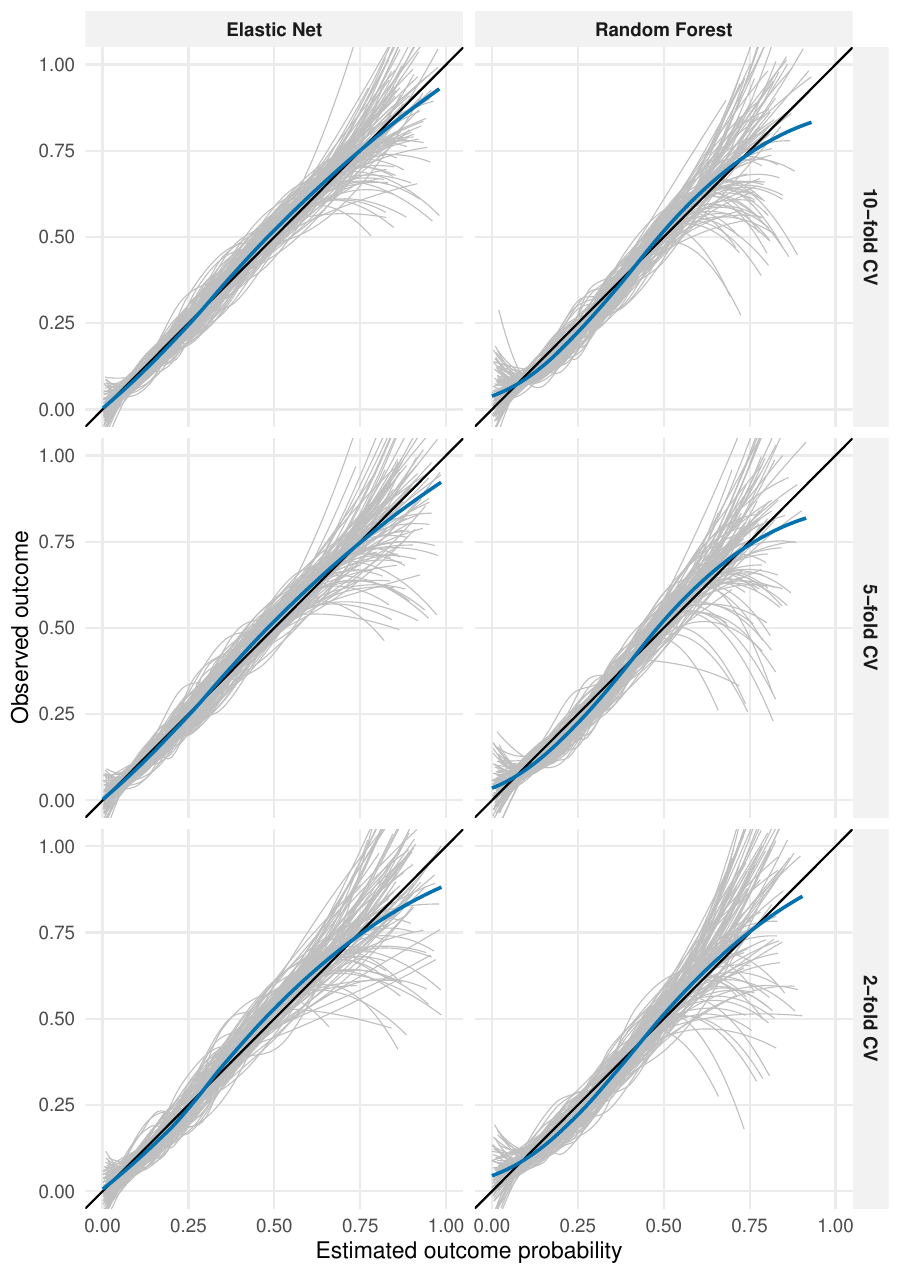}

}

\caption{\label{sfig-calibration-cv-lin}Calibration of the models fitted
on the first 10-, 5- and 2-fold cross-validation samples for the linear
DGM. Thin grey lines show initial dataset-specific LOESS curves; blue
lines show pooled LOESS curves within each scenario and model.}

\end{suppfig}%

\begin{suppfig}

\centering{

\includegraphics[width=0.8\linewidth,height=\textheight,keepaspectratio]{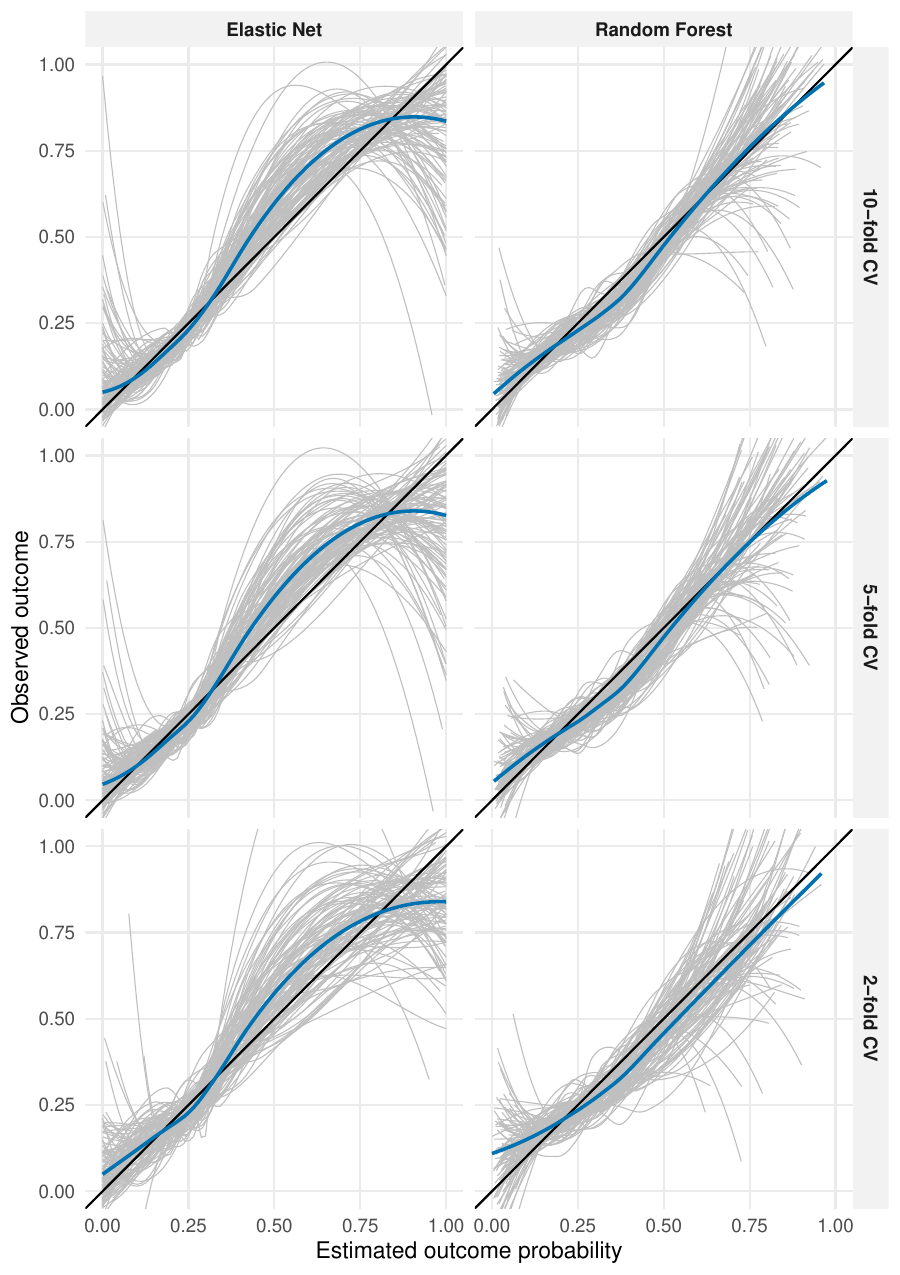}

}

\caption{\label{sfig-calibration-cv-nonlin}Calibration of the models
fitted on the first 10-, 5- and 2-fold cross-validation samples for the
non-linear DGM. Thin grey lines show initial dataset-specific LOESS
curves; blue lines show pooled LOESS curves within each scenario and
model.}

\end{suppfig}%

\FloatBarrier

\begin{suppfig}

\centering{

\pandocbounded{\includegraphics[keepaspectratio]{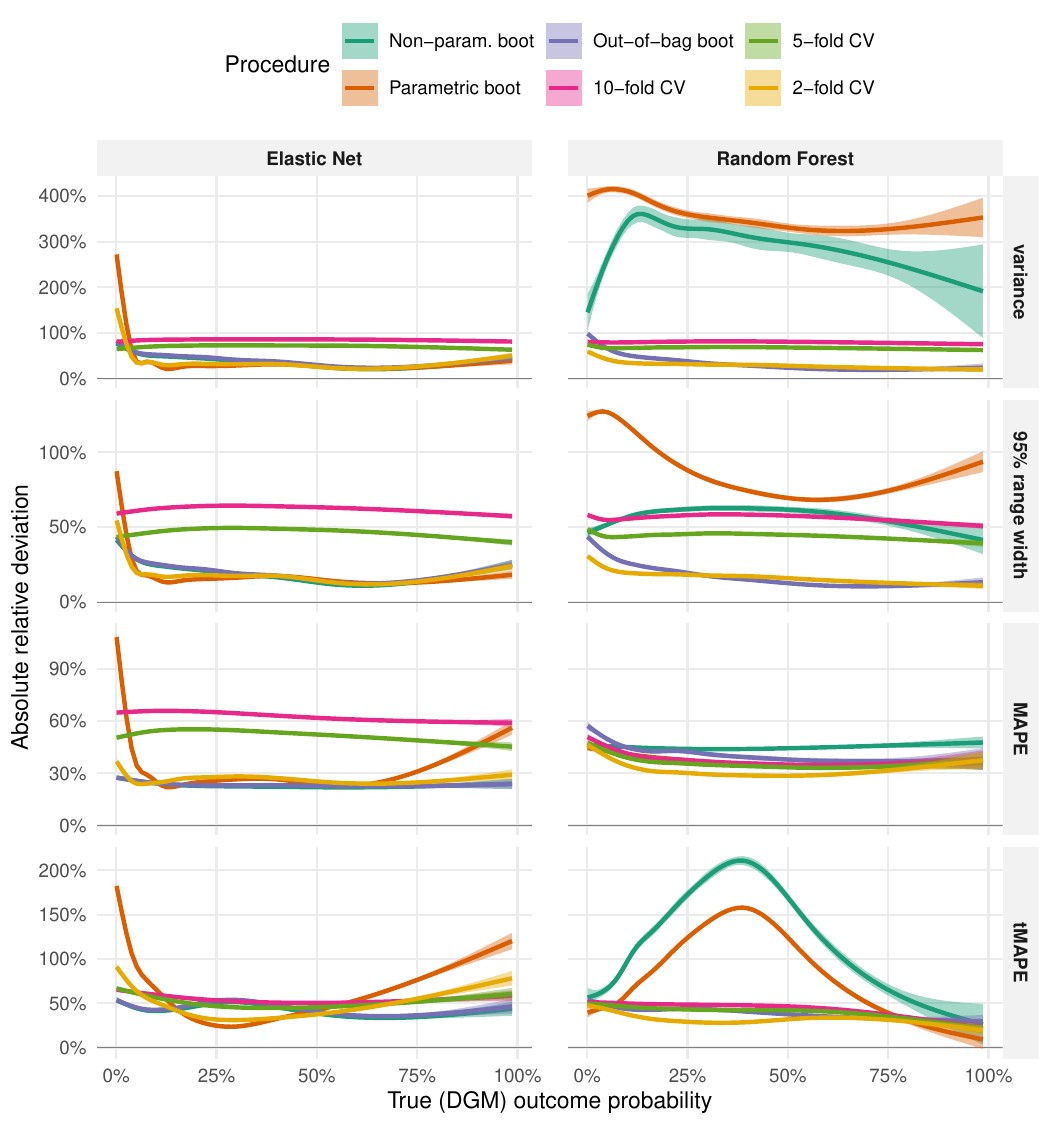}}

}

\caption{\label{sfig-risk-patterns-smaller}Absolute relative
individual-level deviation from independent sampling across the true
(DGM) outcome probability range in the smaller sample-size scenario for
the linear DGM. Each individual-level absolute difference was divided by
the corresponding independent-sampling reference value. Lines are
generalised additive model estimates of the conditional mean, fitted
separately by resampling procedure and model; shaded bands show 95\%
confidence intervals.}

\end{suppfig}%

\begin{suppfig}

\centering{

\pandocbounded{\includegraphics[keepaspectratio]{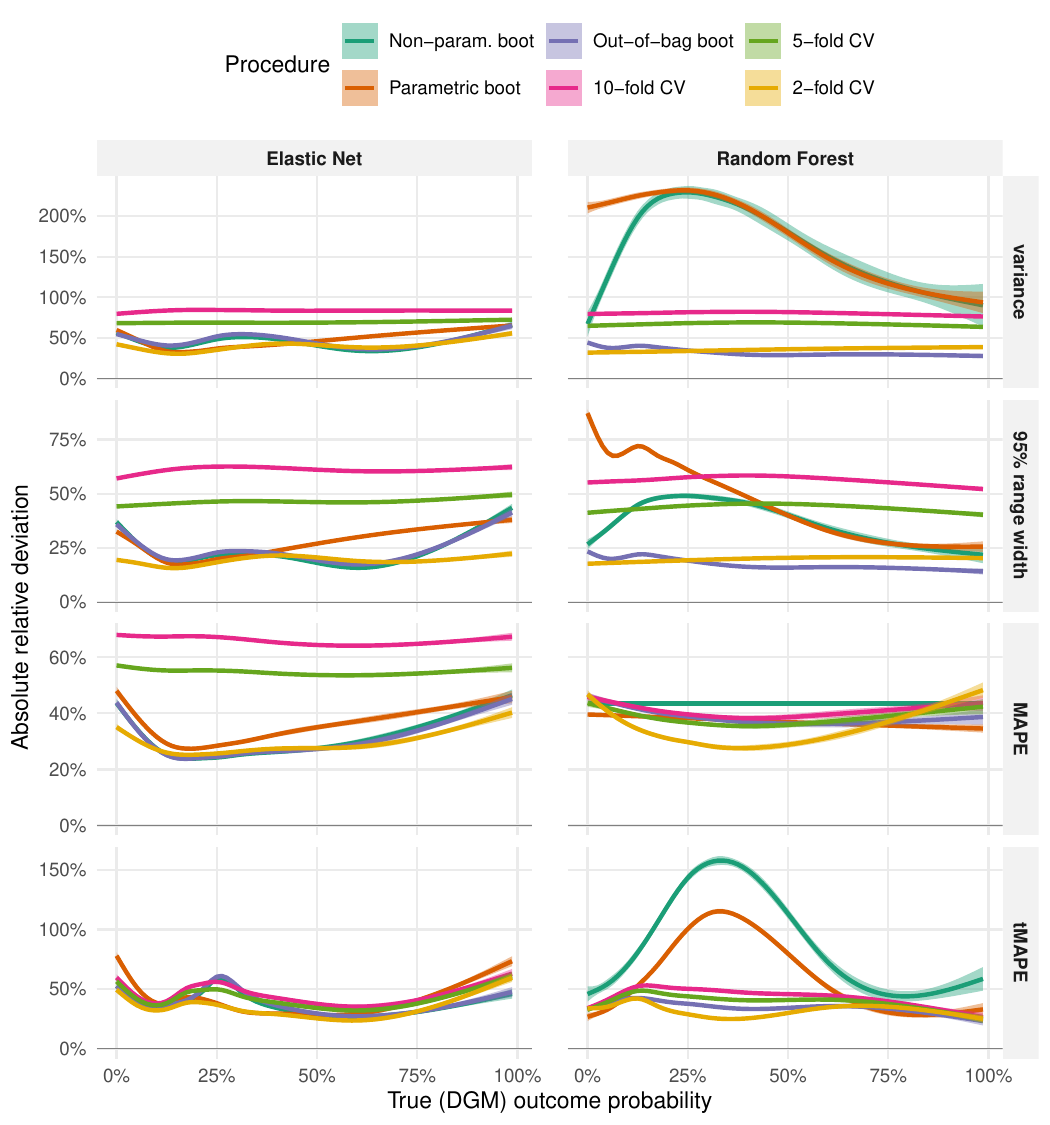}}

}

\caption{\label{sfig-risk-patterns-smaller-nonlin}Absolute relative
individual-level deviation from independent sampling across the true
(DGM) outcome probability range in the smaller sample-size scenario for
the non-linear DGM. Each individual-level absolute difference was
divided by the corresponding independent-sampling reference value. Lines
are generalised additive model estimates of the conditional mean, fitted
separately by resampling procedure and model; shaded bands show 95\%
confidence intervals.}

\end{suppfig}%

\begin{suppfig}

\centering{

\pandocbounded{\includegraphics[keepaspectratio]{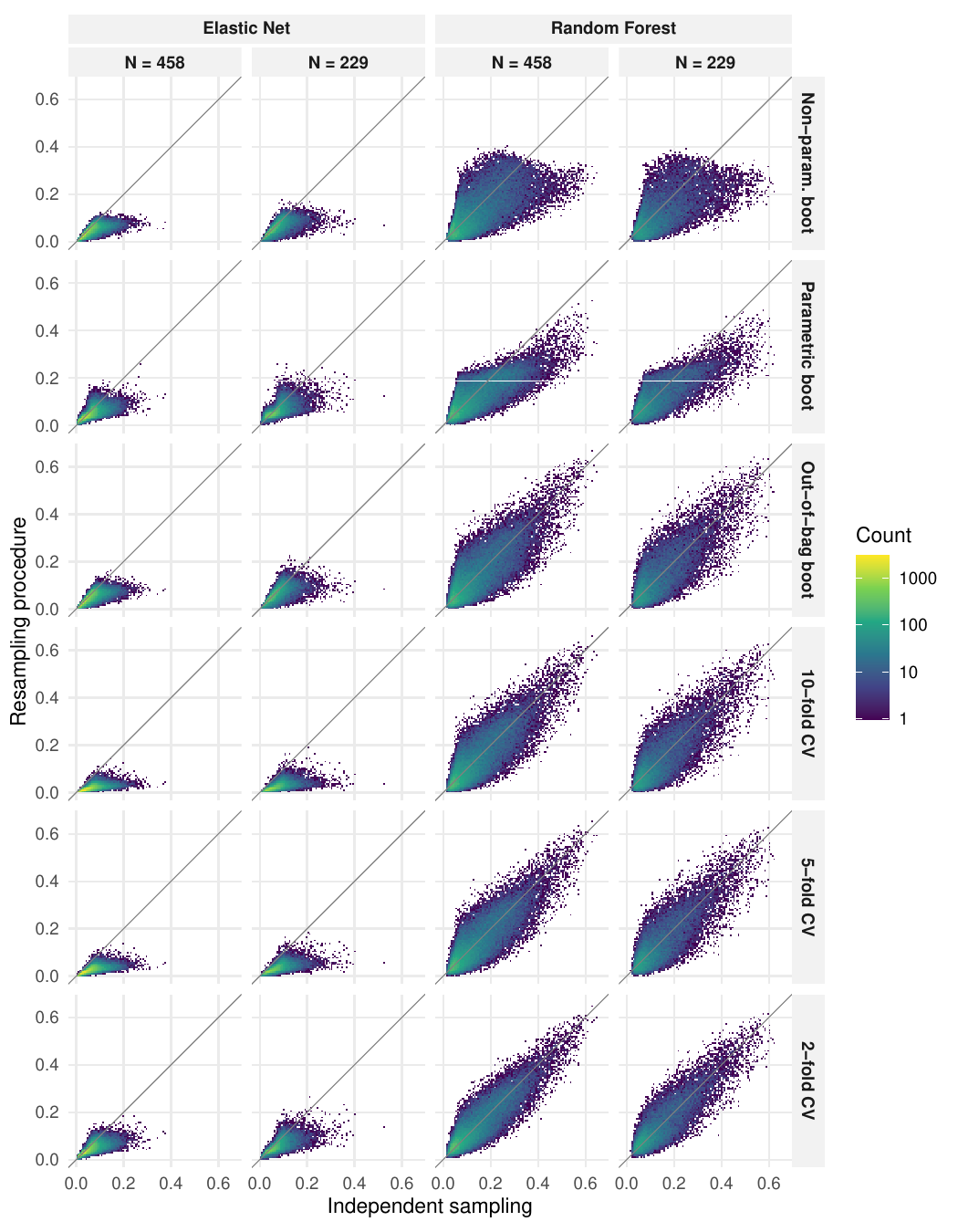}}

}

\caption{\label{sfig-agreement-mape}Individual-level agreement with
independent sampling for MAPE under the linear DGM.}

\end{suppfig}%

\begin{suppfig}

\centering{

\pandocbounded{\includegraphics[keepaspectratio]{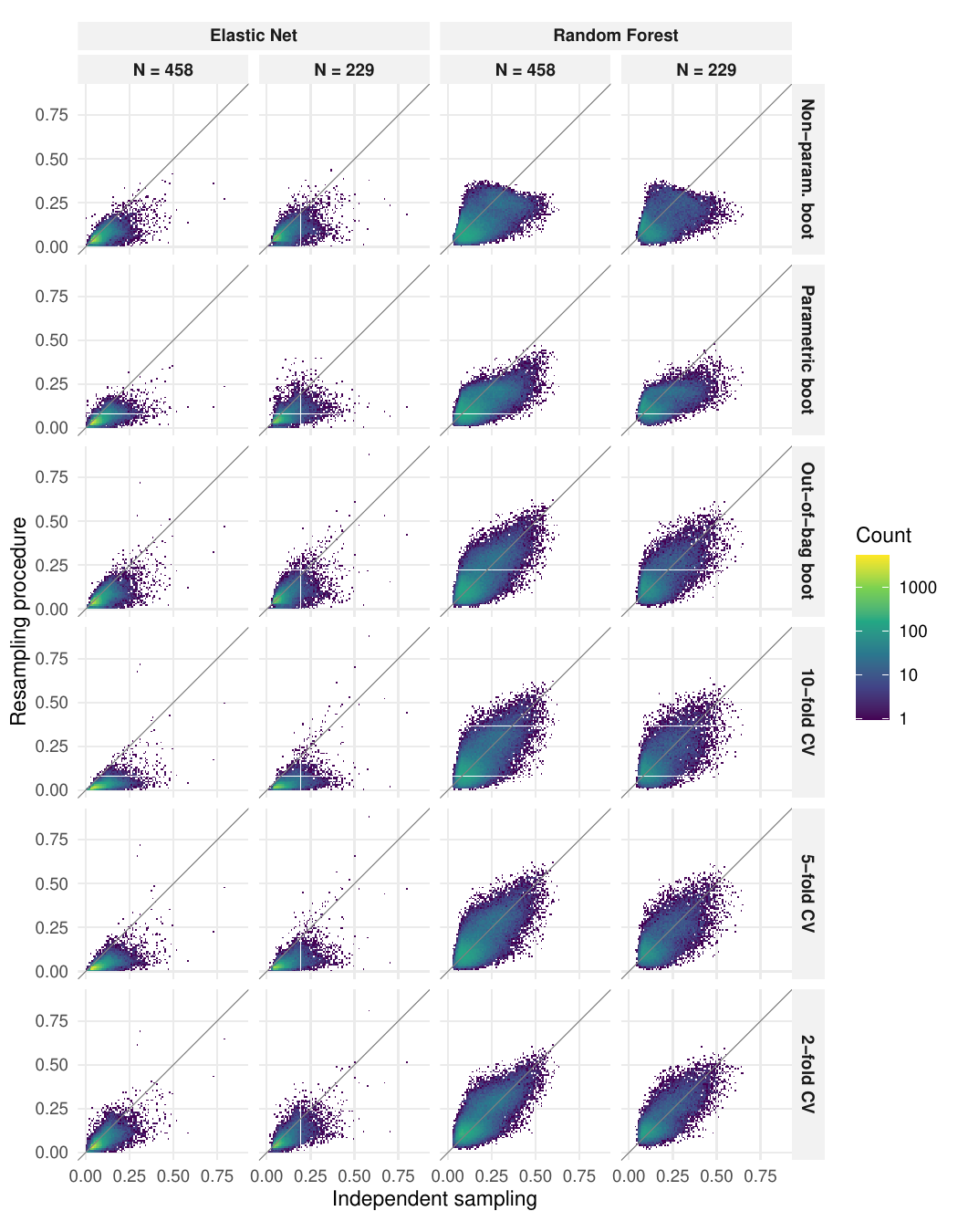}}

}

\caption{\label{sfig-agreement-mape-nonlin}Individual-level agreement
with independent sampling for MAPE under the non-linear DGM.}

\end{suppfig}%

\begin{suppfig}

\centering{

\pandocbounded{\includegraphics[keepaspectratio]{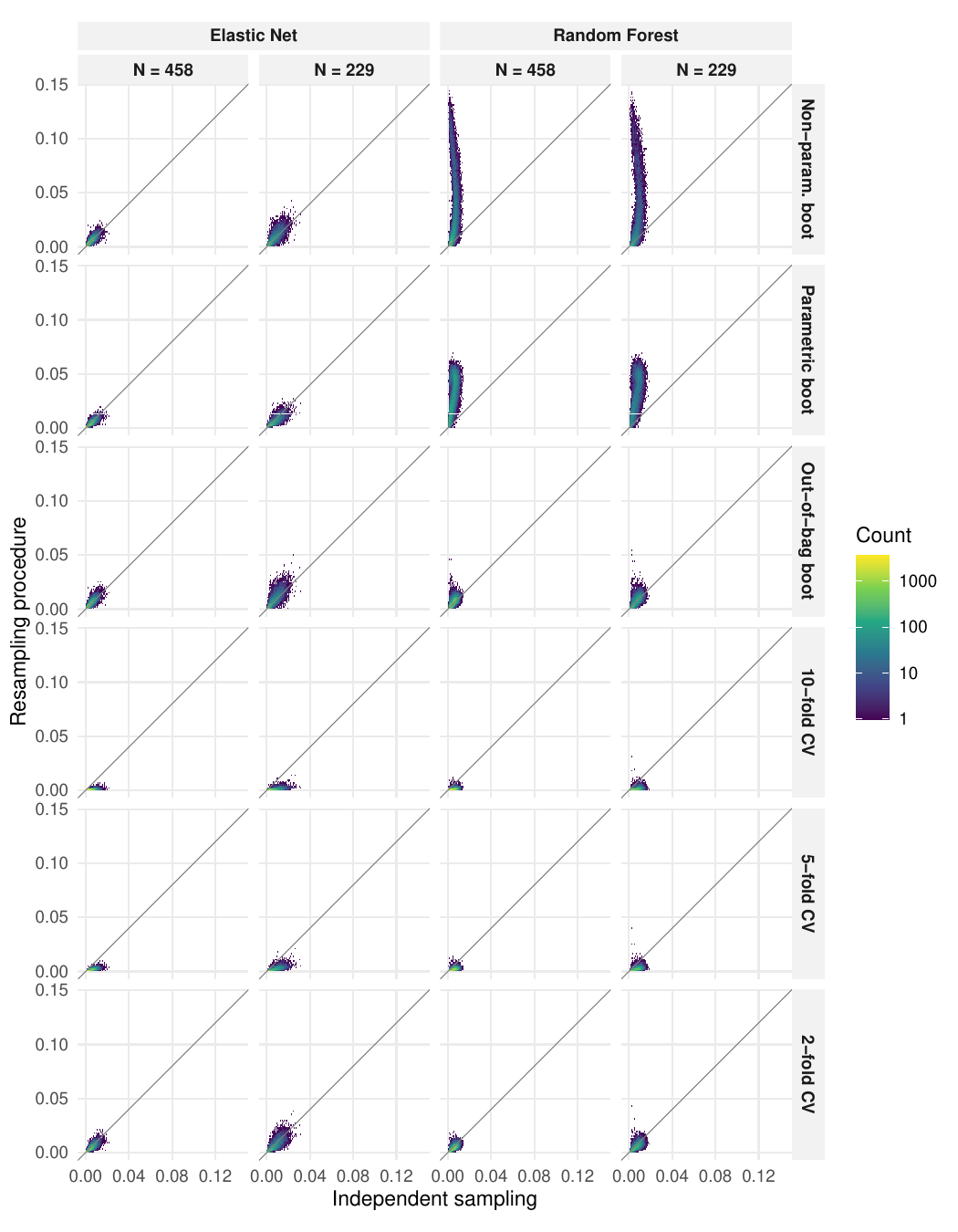}}

}

\caption{\label{sfig-agreement-variance}Individual-level agreement with
independent sampling for prediction variance under the linear DGM.}

\end{suppfig}%

\begin{suppfig}

\centering{

\pandocbounded{\includegraphics[keepaspectratio]{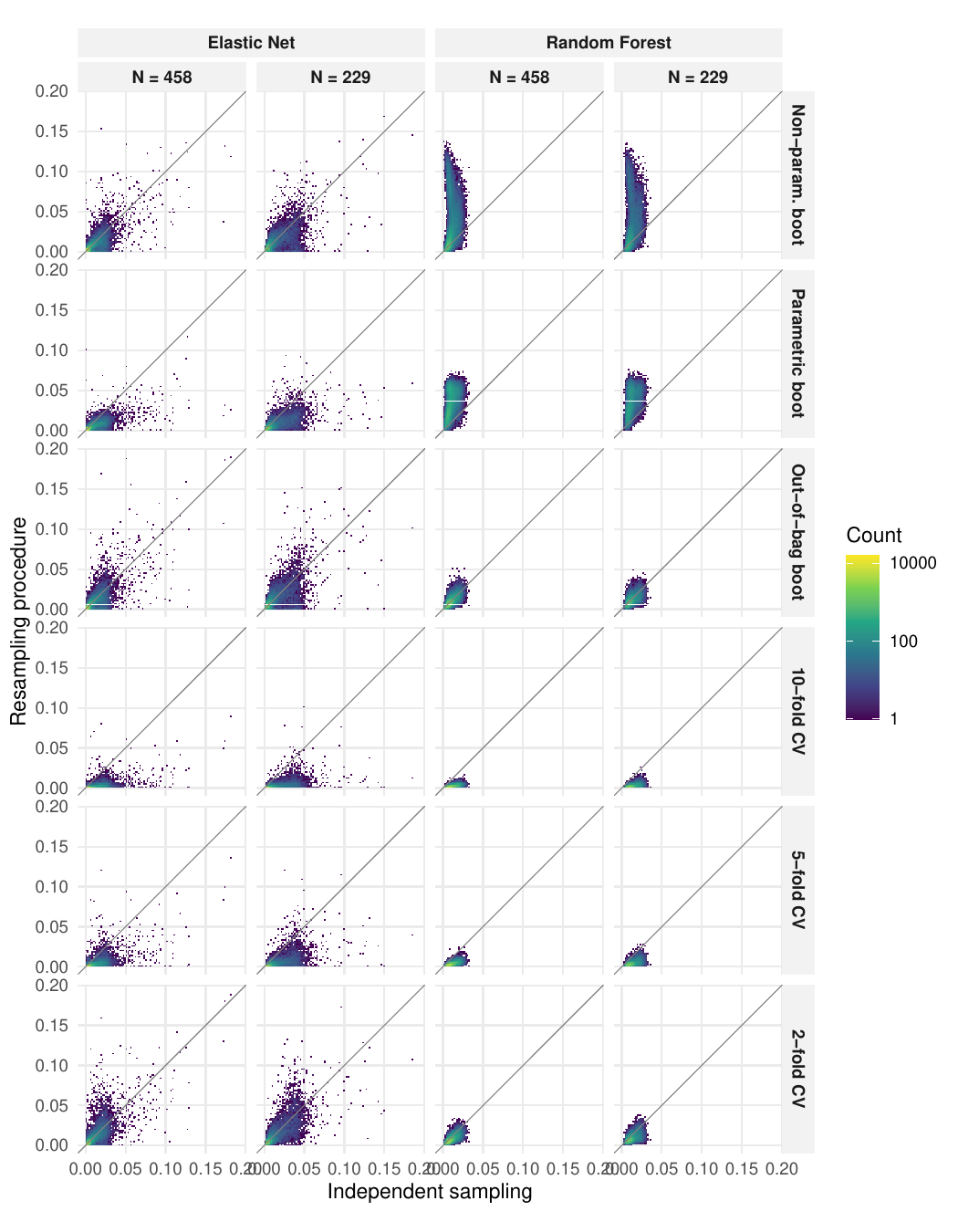}}

}

\caption{\label{sfig-agreement-variance-nonlin}Individual-level
agreement with independent sampling for prediction variance under the
non-linear DGM.}

\end{suppfig}%

\begin{suppfig}

\centering{

\pandocbounded{\includegraphics[keepaspectratio]{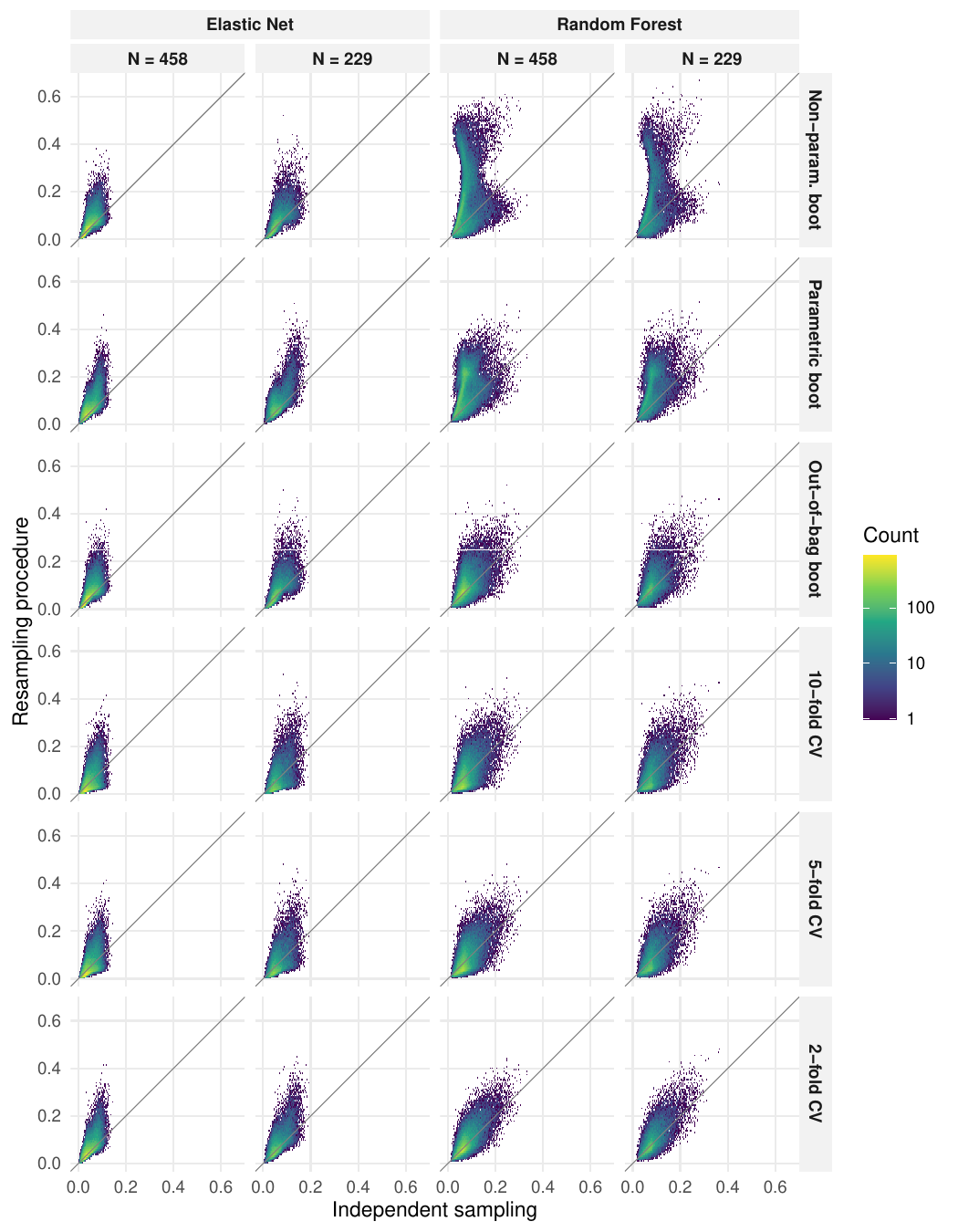}}

}

\caption{\label{sfig-agreement-tmape}Individual-level agreement with
independent sampling for tMAPE under the linear DGM.}

\end{suppfig}%

\begin{suppfig}

\centering{

\pandocbounded{\includegraphics[keepaspectratio]{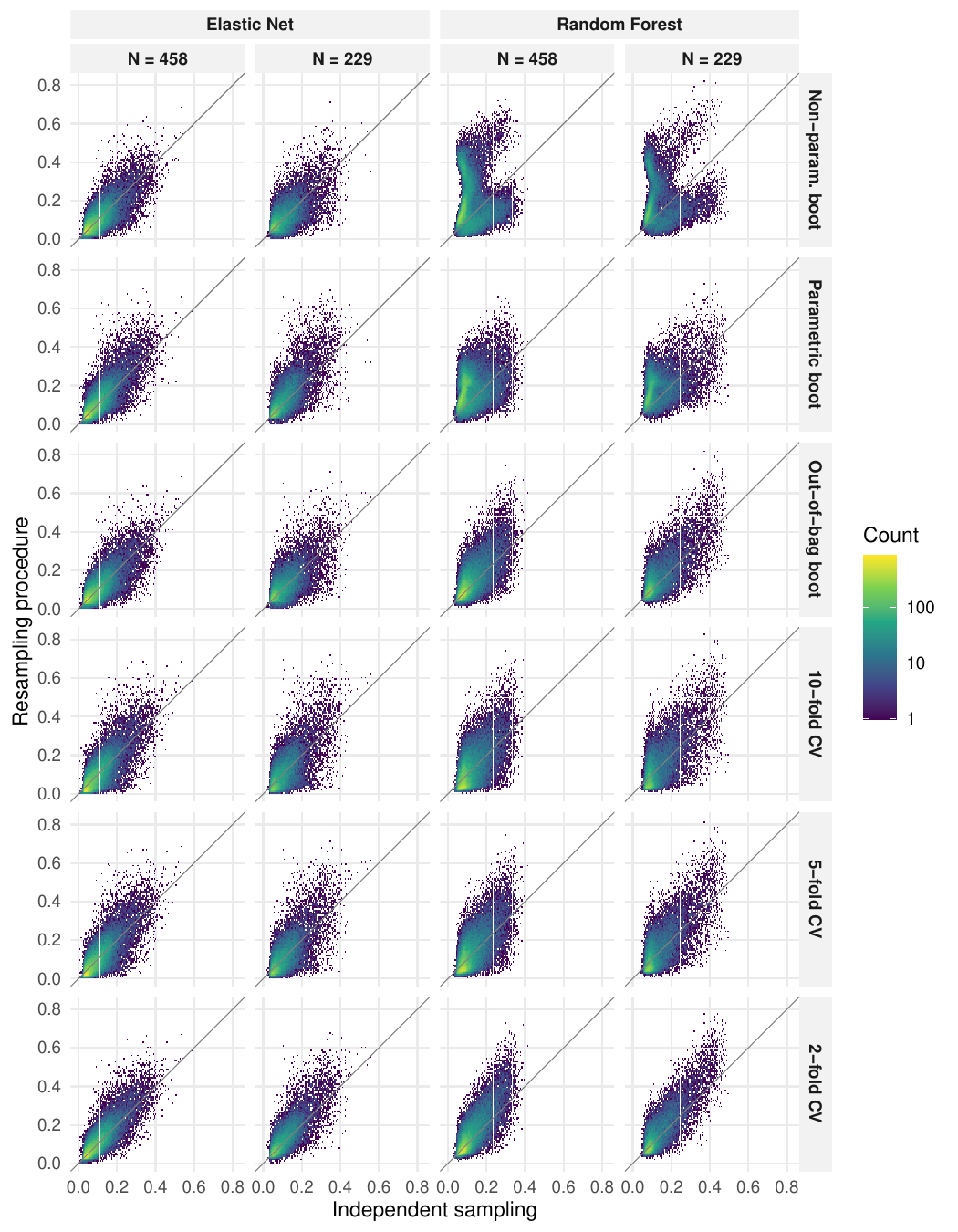}}

}

\caption{\label{sfig-agreement-tmape-nonlin}Individual-level agreement
with independent sampling for tMAPE under the non-linear DGM.}

\end{suppfig}%

\begin{suppfig}

\centering{

\pandocbounded{\includegraphics[keepaspectratio]{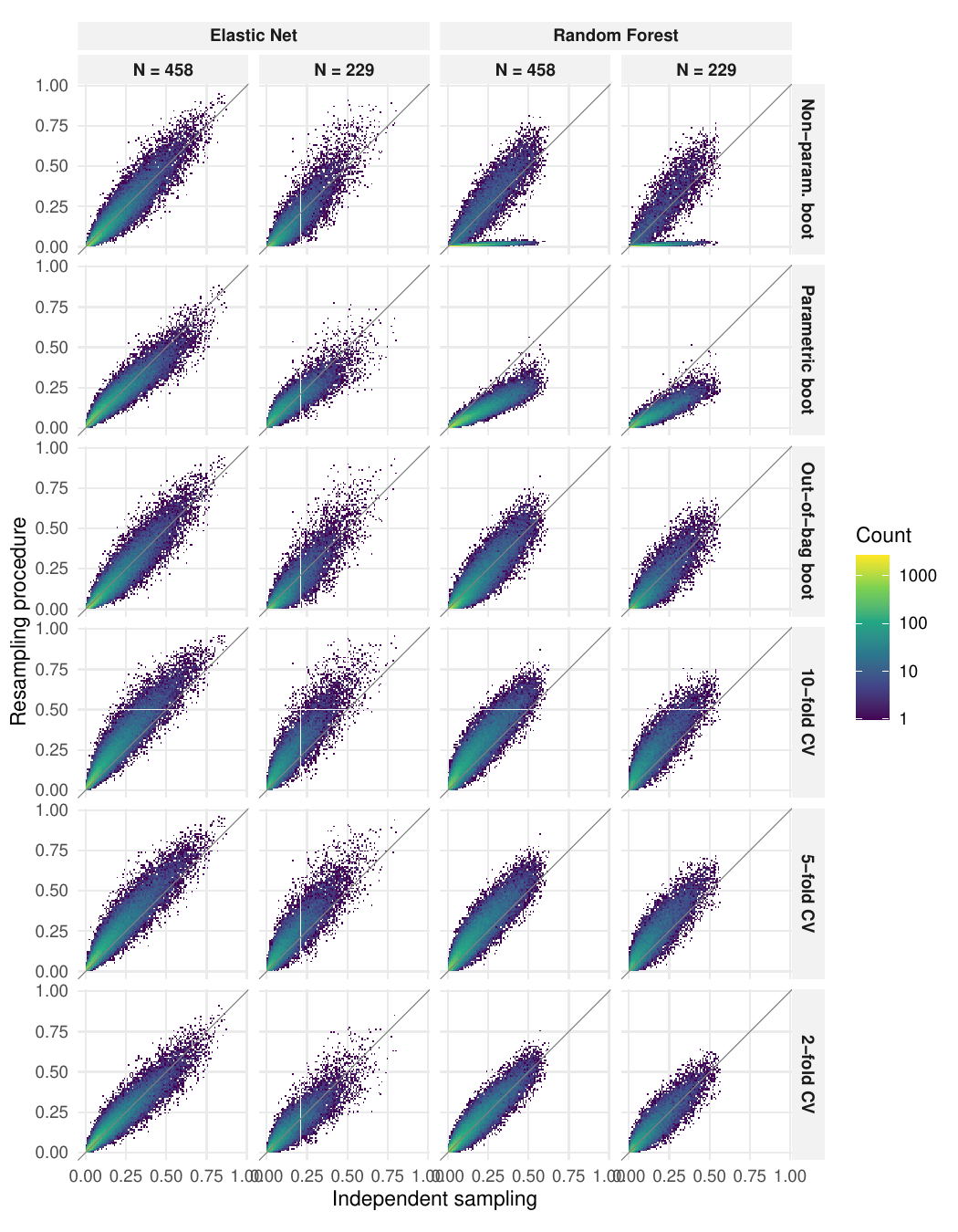}}

}

\caption{\label{sfig-agreement-q025}Individual-level agreement with
independent sampling for the 2.5th percentile under the linear DGM.}

\end{suppfig}%

\begin{suppfig}

\centering{

\pandocbounded{\includegraphics[keepaspectratio]{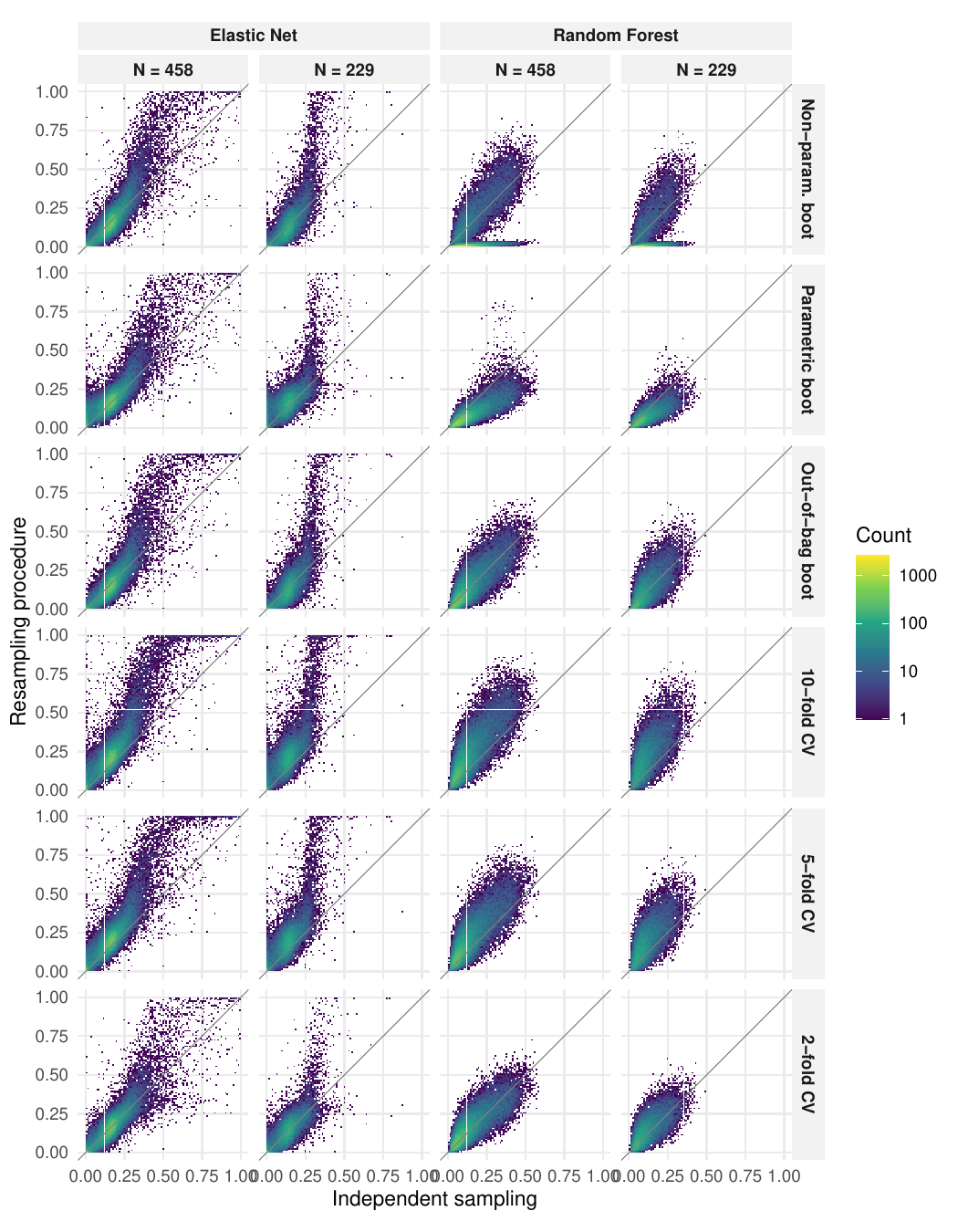}}

}

\caption{\label{sfig-agreement-q025-nonlin}Individual-level agreement
with independent sampling for the 2.5th percentile under the non-linear
DGM.}

\end{suppfig}%

\begin{suppfig}

\centering{

\pandocbounded{\includegraphics[keepaspectratio]{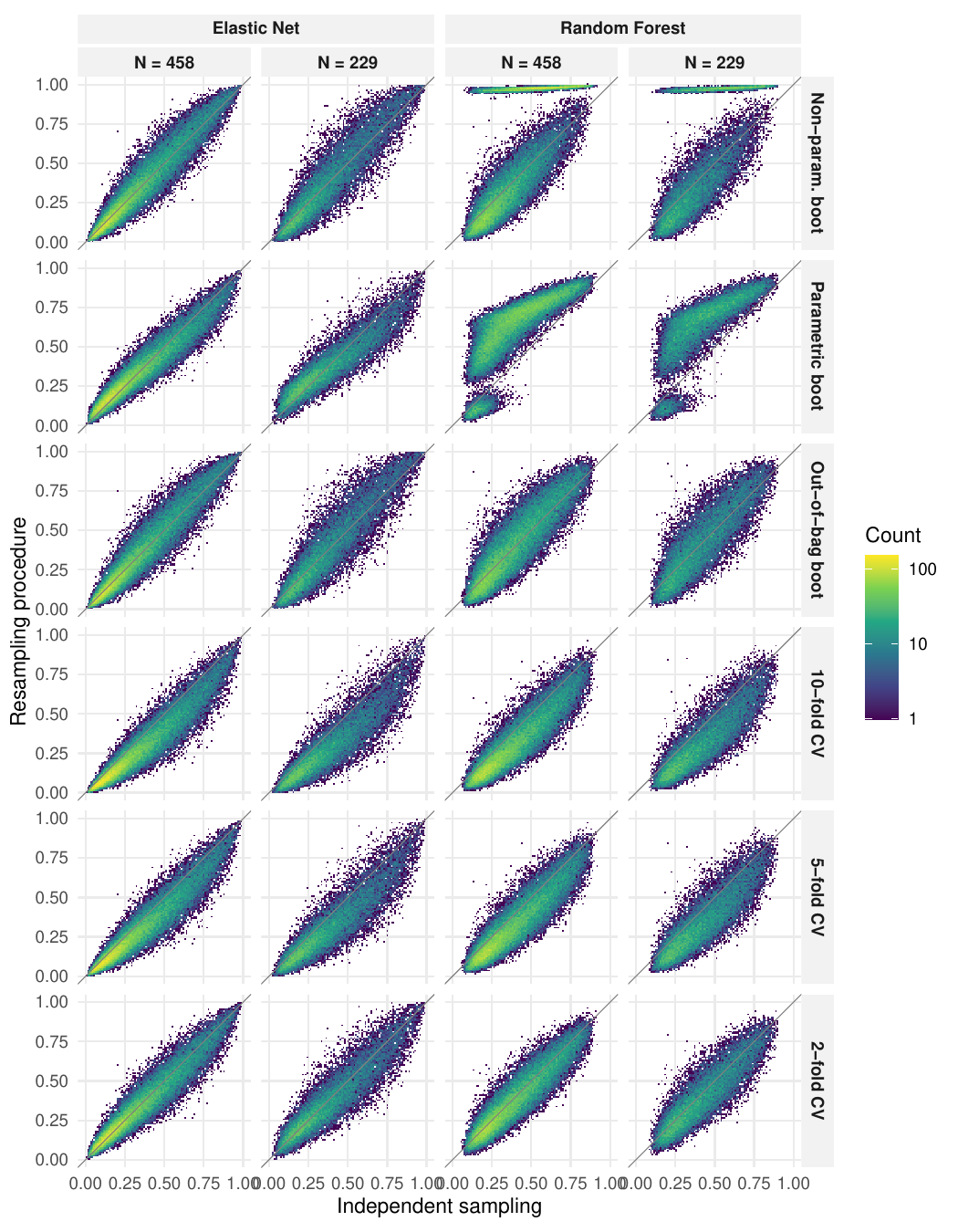}}

}

\caption{\label{sfig-agreement-q975}Individual-level agreement with
independent sampling for the 97.5th percentile under the linear DGM.}

\end{suppfig}%

\begin{suppfig}

\centering{

\pandocbounded{\includegraphics[keepaspectratio]{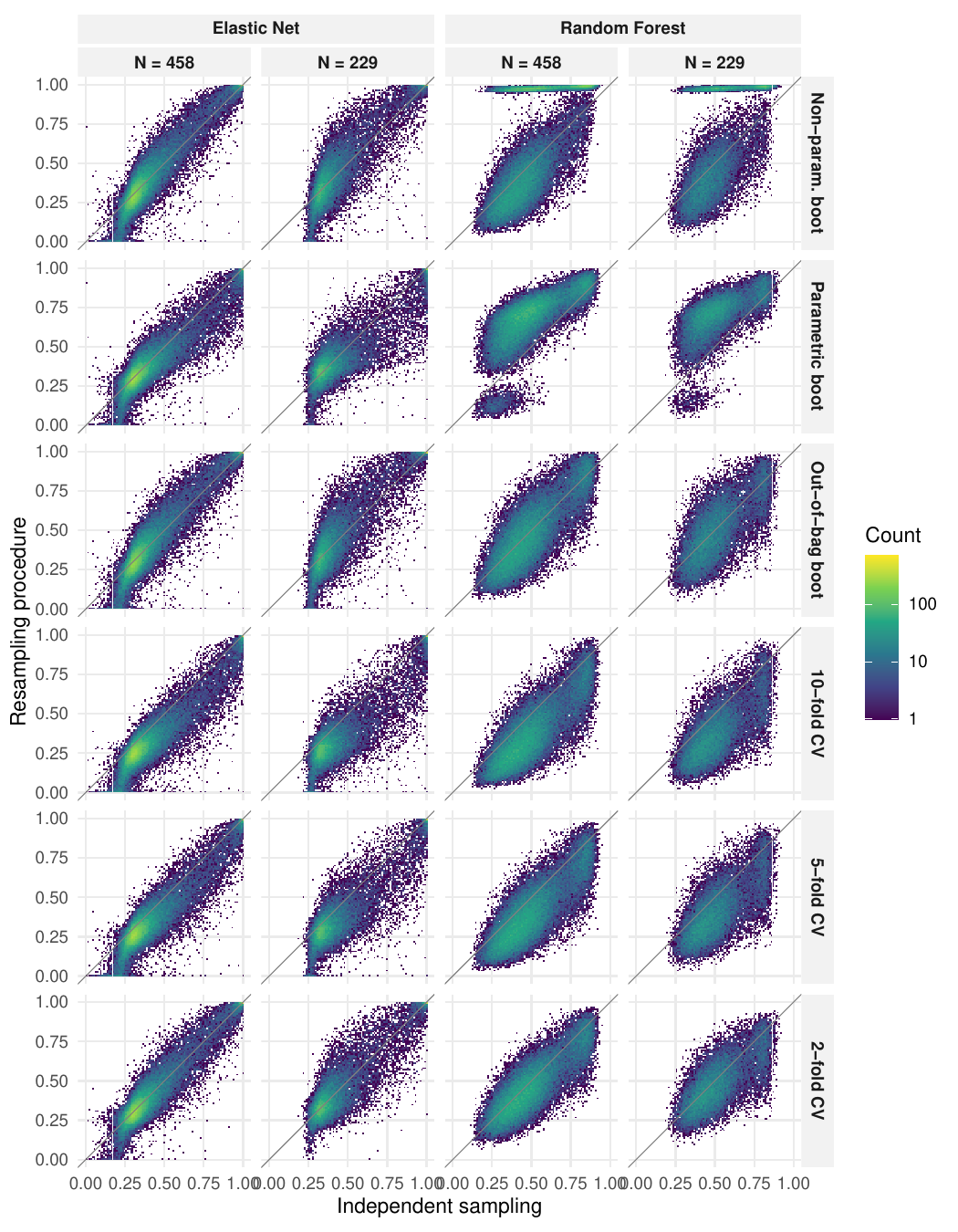}}

}

\caption{\label{sfig-agreement-q975-nonlin}Individual-level agreement
with independent sampling for the 97.5th percentile under the non-linear
DGM.}

\end{suppfig}%

\begin{landscape}

\begin{suppfig}

\centering{

\pandocbounded{\includegraphics[keepaspectratio]{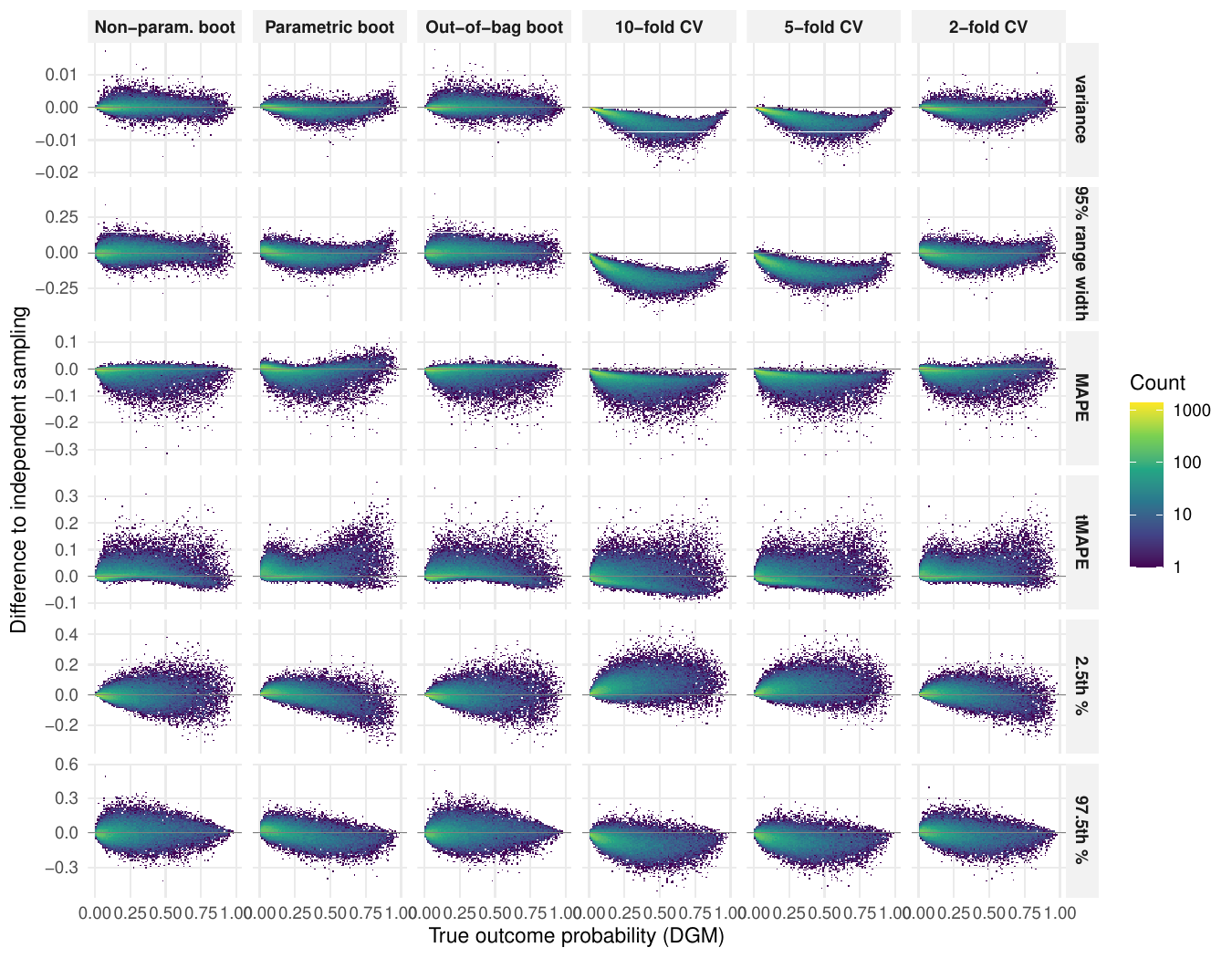}}

}

\caption{\label{sfig-original-risk-patterns-glmnet}Difference from
independent sampling over the true DGM probability range for the Elastic
Net model under the linear DGM in the larger sample-size scenario.}

\end{suppfig}%

\end{landscape}

\begin{landscape}

\begin{suppfig}

\centering{

\pandocbounded{\includegraphics[keepaspectratio]{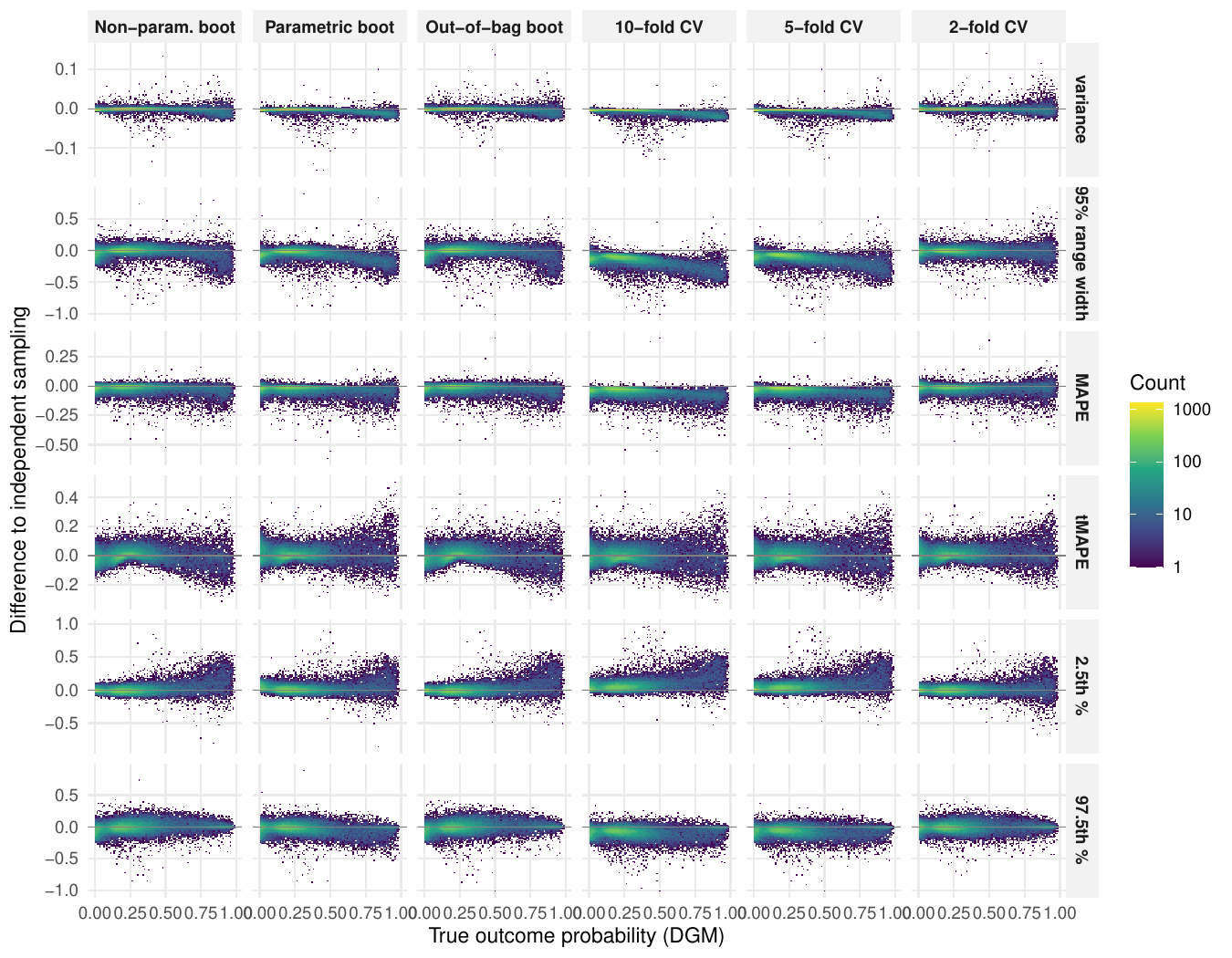}}

}

\caption{\label{sfig-original-risk-patterns-glmnet-nonlin}Difference
from independent sampling over the true DGM probability range for the
Elastic Net model under the non-linear DGM in the larger sample-size
scenario.}

\end{suppfig}%

\end{landscape}

\begin{landscape}

\begin{suppfig}

\centering{

\pandocbounded{\includegraphics[keepaspectratio]{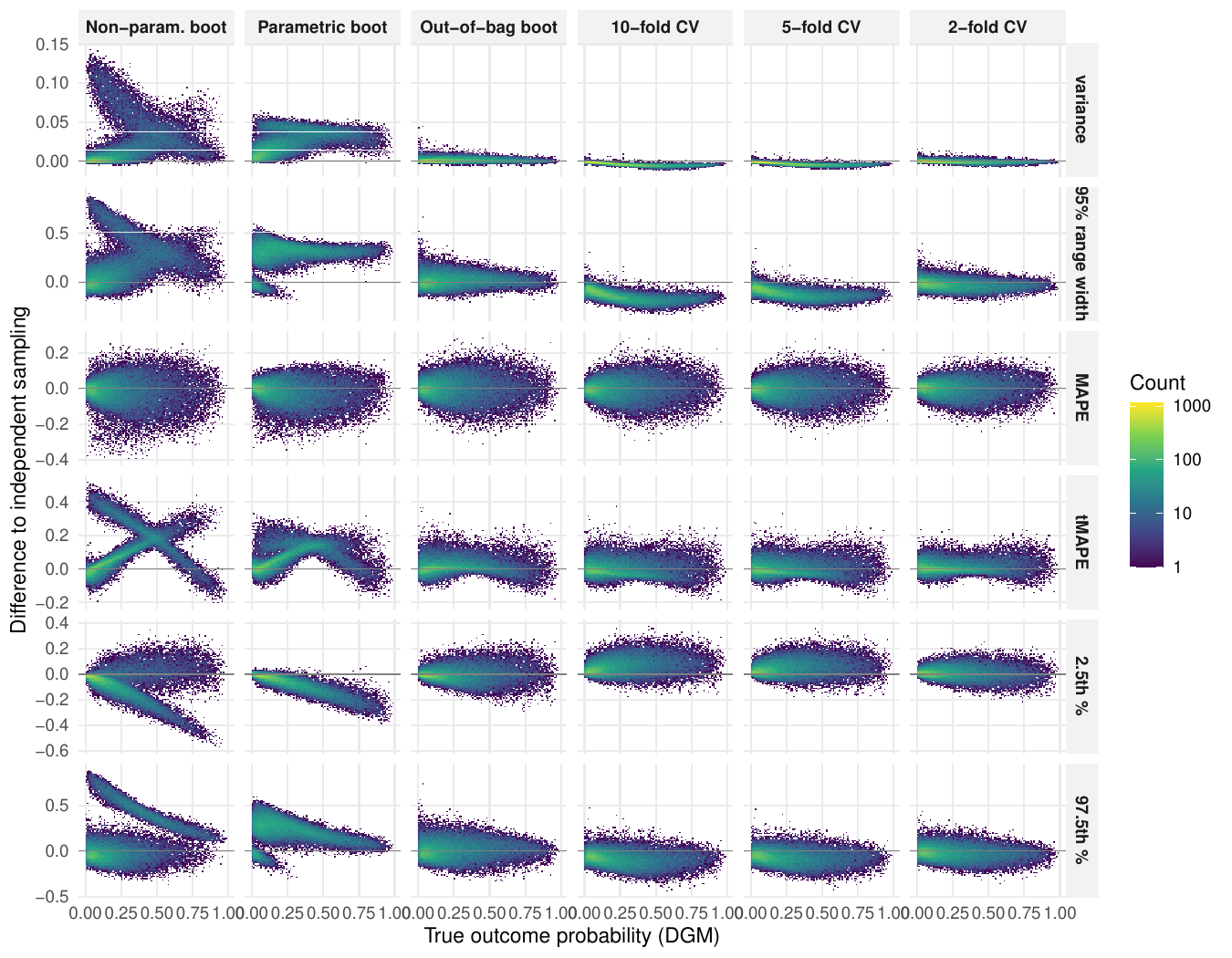}}

}

\caption{\label{sfig-original-risk-patterns-rf}Difference from
independent sampling over the true DGM probability range for the Random
Forest model under the linear DGM in the larger sample-size scenario.}

\end{suppfig}%

\end{landscape}

\begin{landscape}

\begin{suppfig}

\centering{

\pandocbounded{\includegraphics[keepaspectratio]{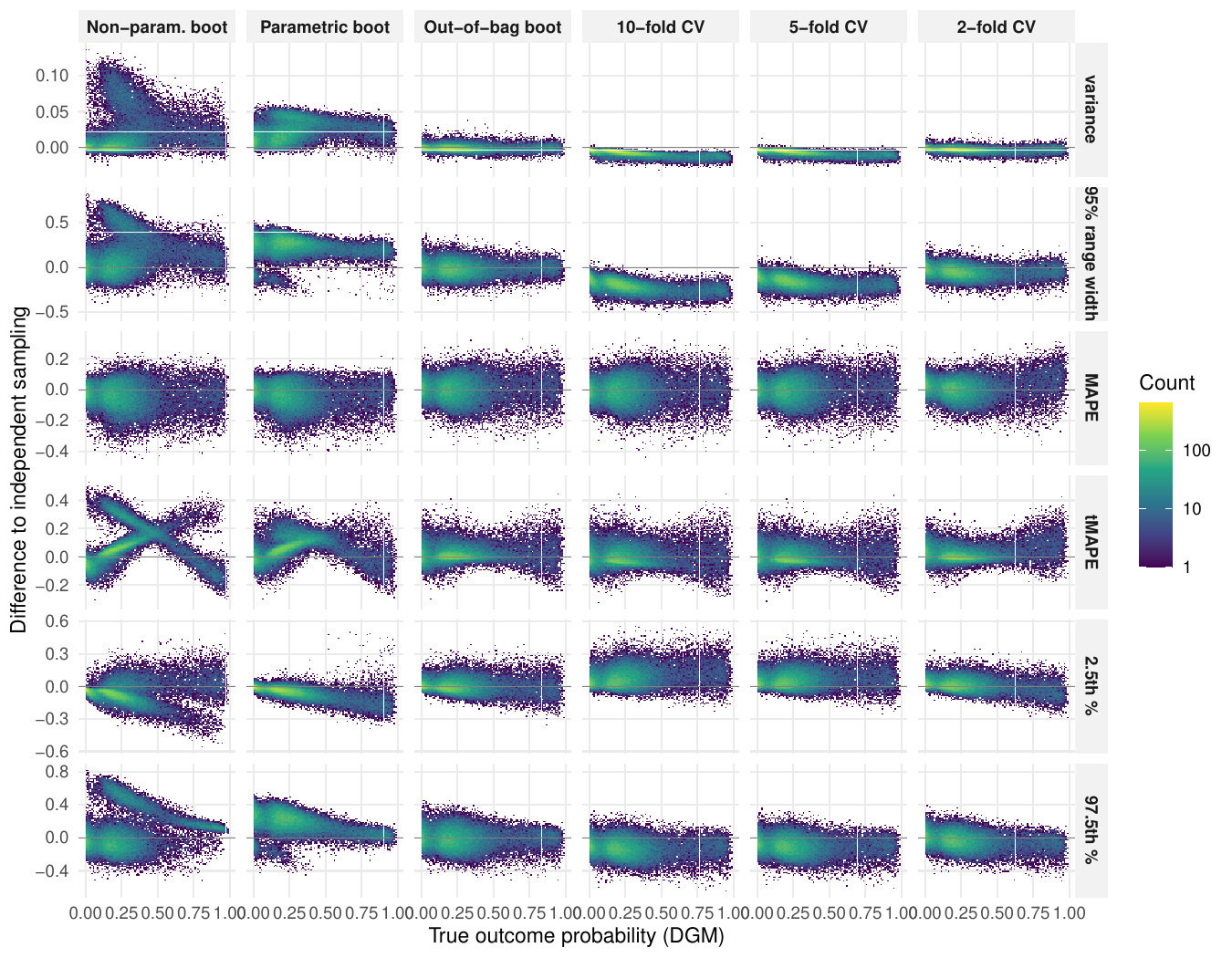}}

}

\caption{\label{sfig-original-risk-patterns-rf-nonlin}Difference from
independent sampling over the true DGM probability range for the Random
Forest model under the non-linear DGM in the larger sample-size
scenario.}

\end{suppfig}%

\end{landscape}

\FloatBarrier

\newpage{}

\subsection{Original-fit hyperparameters}\label{sec-hyperpars}

Only hyperparameters from the models fitted to the original datasets
were saved. Their joint and marginal distributions are shown for Elastic
Net under the linear and non-linear DGMs
(S-Figure~\ref{sfig-hyperparameters-glmnet};
S-Figure~\ref{sfig-hyperparameters-glmnet-nonlin}) and for Random Forest
under both DGMs (S-Figure~\ref{sfig-hyperparameters-rf};
S-Figure~\ref{sfig-hyperparameters-rf-nonlin}).

\begin{suppfig}

\centering{

\pandocbounded{\includegraphics[keepaspectratio]{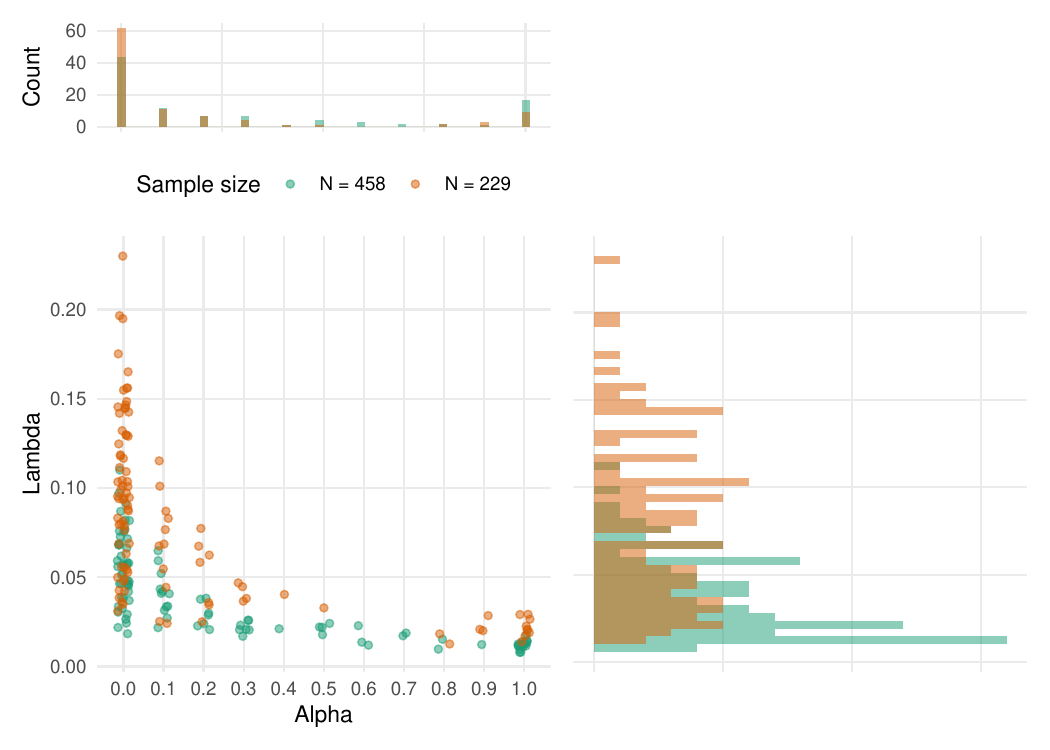}}

}

\caption{\label{sfig-hyperparameters-glmnet}Joint and marginal
distributions of the hyperparameters selected for the original Elastic
Net fits under the linear DGM.}

\end{suppfig}%

\begin{suppfig}

\centering{

\pandocbounded{\includegraphics[keepaspectratio]{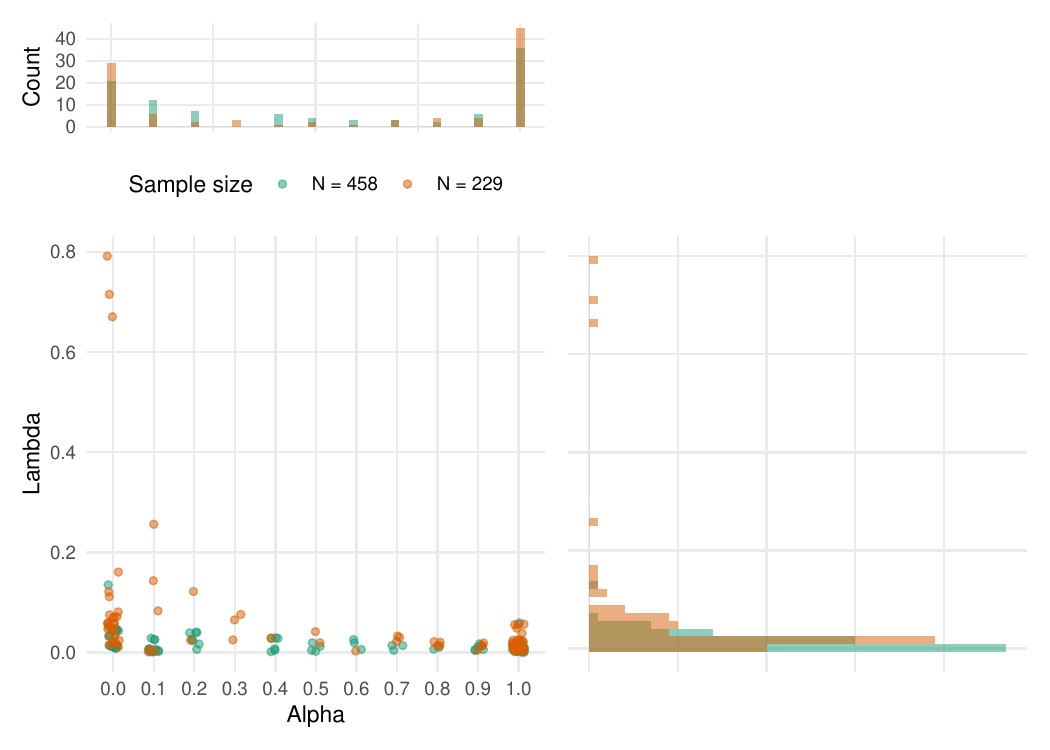}}

}

\caption{\label{sfig-hyperparameters-glmnet-nonlin}Joint and marginal
distributions of the hyperparameters selected for the original Elastic
Net fits under the non-linear DGM.}

\end{suppfig}%

\begin{suppfig}

\centering{

\pandocbounded{\includegraphics[keepaspectratio]{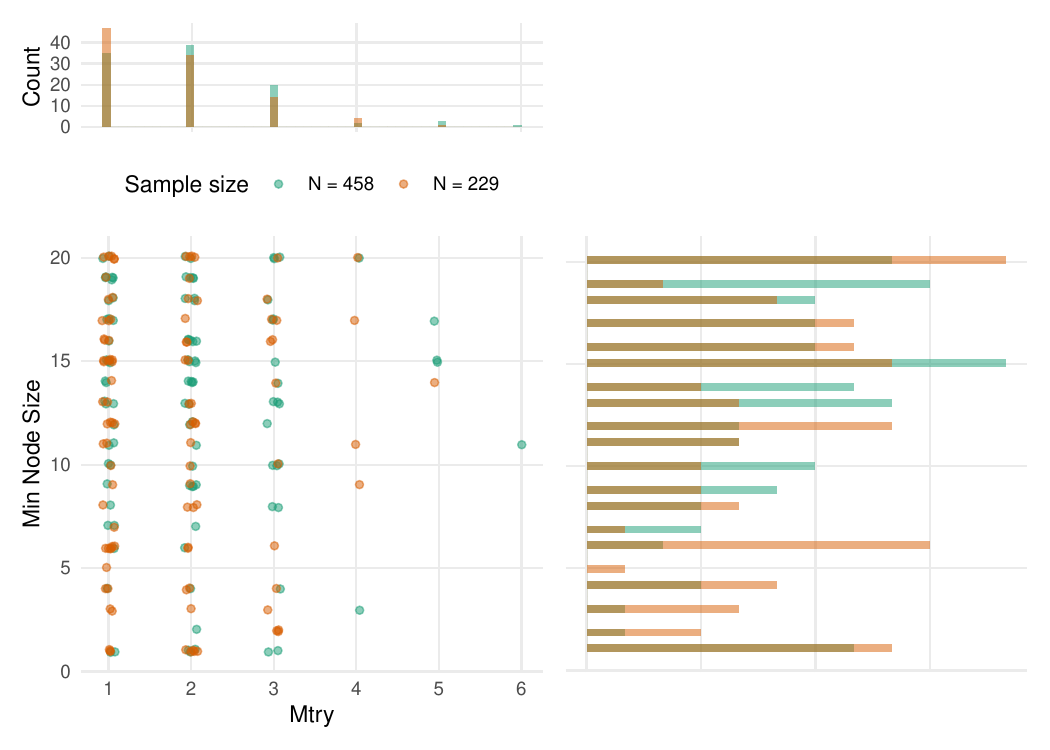}}

}

\caption{\label{sfig-hyperparameters-rf}Joint and marginal distributions
of the hyperparameters selected for the original Random Forest fits
under the linear DGM.}

\end{suppfig}%

\begin{suppfig}

\centering{

\pandocbounded{\includegraphics[keepaspectratio]{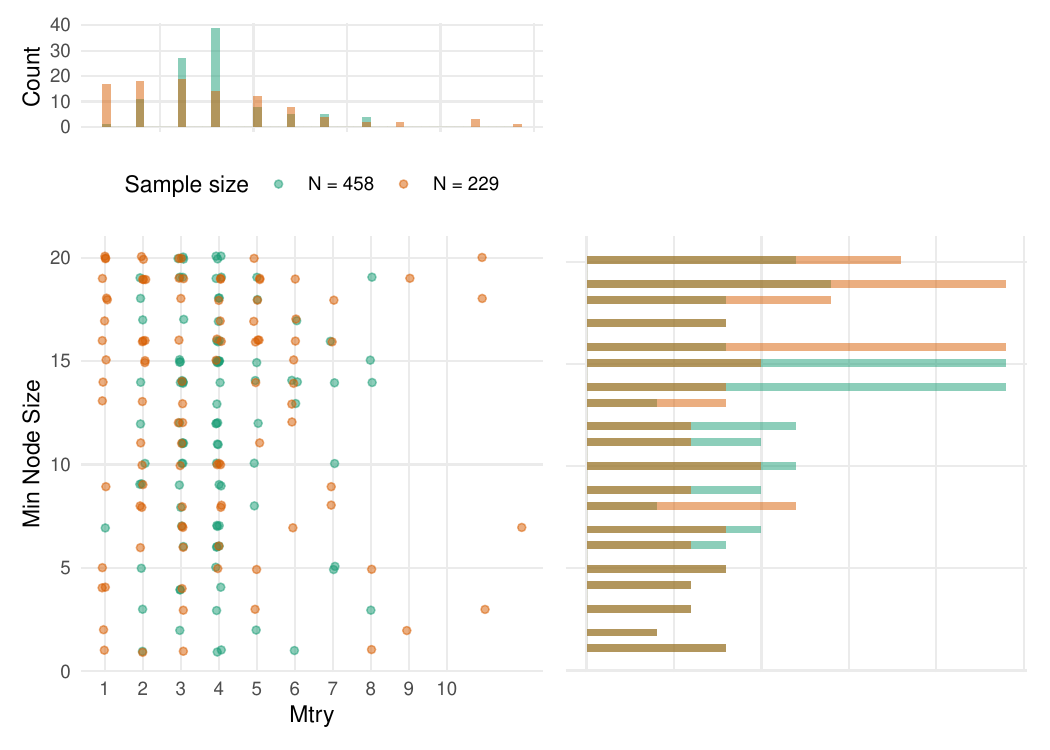}}

}

\caption{\label{sfig-hyperparameters-rf-nonlin}Joint and marginal
distributions of the hyperparameters selected for the original Random
Forest fits under the non-linear DGM.}

\end{suppfig}%

\FloatBarrier

\subsubsection{An influential observation in the outcome
probability-pattern curves}\label{sec-risk-pattern-outlier}

One outlying observation was removed from
Figure~\ref{fig-results-risk-patterns-nonlin}. This observation caused
the confidence bands for the non-parametric and default parametric
bootstrap in the Elastic Net empirical 95\% range-width panel to be very
wide.

S-Figure~\ref{sfig-results-risk-patterns-nonlin-outlier} shows the
figure with the observation included.

\begin{suppfig}

\centering{

\pandocbounded{\includegraphics[keepaspectratio]{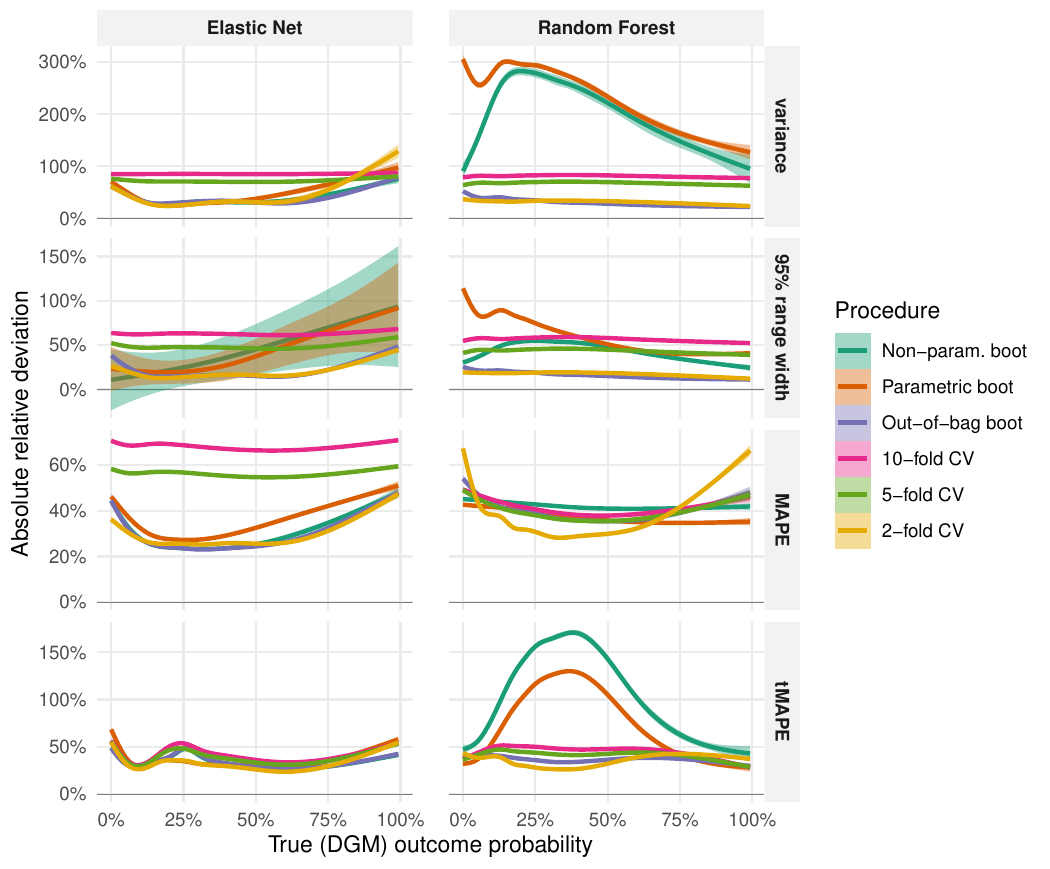}}

}

\caption{\label{sfig-results-risk-patterns-nonlin-outlier}Absolute
relative individual-level deviation from independent sampling across the
true (DGM) outcome probability range for the non-linear DGM in the
larger sample-size scenario. Each individual-level absolute difference
was divided by the corresponding independent-sampling reference value.
Lines are generalised additive model estimates of the conditional mean,
fitted separately by resampling procedure and model; shaded bands show
95\% confidence intervals.}

\end{suppfig}%

The excluded subject's true (DGM) outcome probability was 74.72\%,
whereas the original fitted probability was 99.84\%.
S-Figure~\ref{sfig-risk-pattern-outlier} shows the 200 predictions from
each sampling procedure.

Under independent sampling, all but one prediction were very close to
1.0, as were all out-of-bag bootstrap and repeated cross-validation
predictions. Consequently, the 2.5th and 97.5th percentiles were both
close to one for these methods.

In contrast, non-parametric and parametric bootstrap predictions were
more dispersed, with 2.5th percentiles of 26.6\% and 58.2\%,
respectively. Their range widths were 0.734 and 0.418. Dividing the
absolute differences in range width by the almost zero
independent-sampling reference (0.000158) produced enormous relative
deviations, inflating the residual variance of the Gaussian generalised
additive model, widening the confidence bands and influencing the fitted
mean curve.

\begin{suppfig}

\centering{

\pandocbounded{\includegraphics[keepaspectratio]{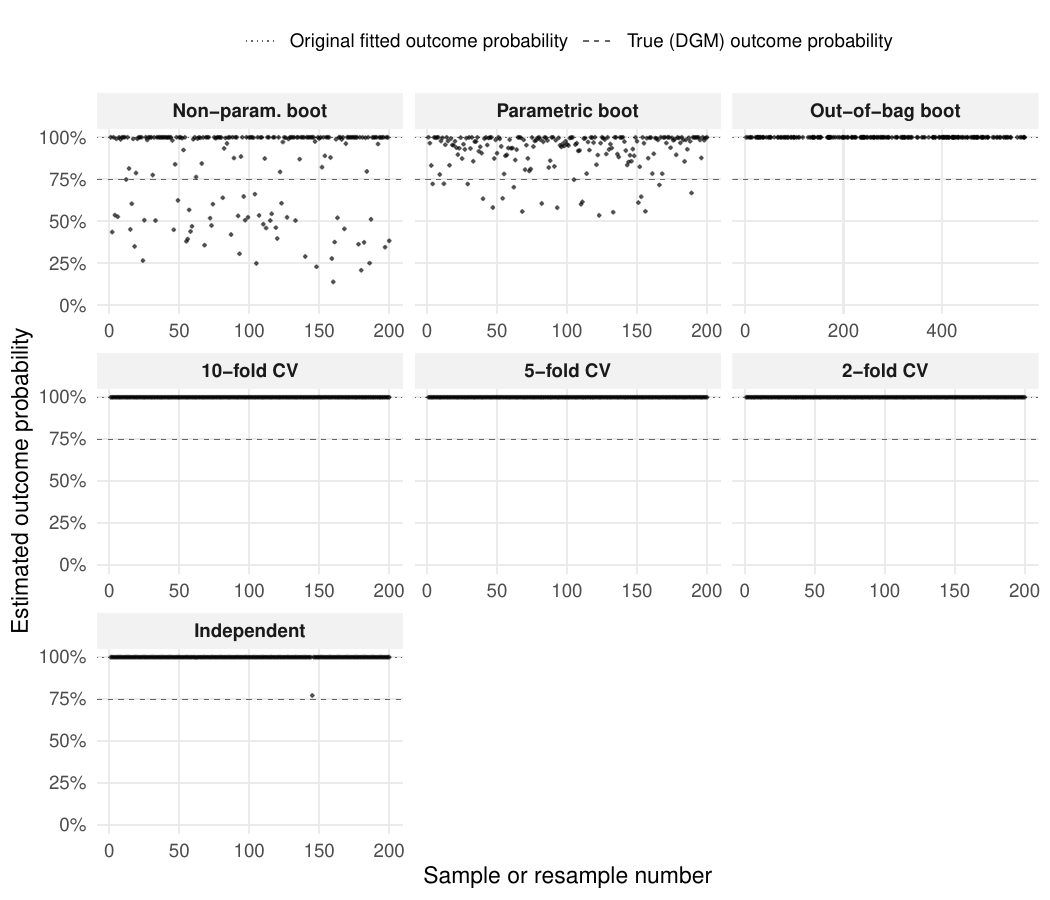}}

}

\caption{\label{sfig-risk-pattern-outlier}Predictions for the
influential individual in the non-linear DGM, larger-sample Elastic Net
scenario. Each point represents one of the 200 predictions under the
indicated sampling procedure. Reference lines show the true (DGM)
outcome probability and the original fitted probability. Predictions
extremely close to one overlap at the top of each panel.}

\end{suppfig}%

\FloatBarrier

\newpage{}

\section*{References}\label{references}
\addcontentsline{toc}{section}{References}

\protect\phantomsection\label{refs}

\end{document}